\documentclass{aa}

\usepackage{graphicx}
\usepackage{txfonts}
\usepackage{color}
\usepackage{longtable}

\newcommand{\tablenotea}[1]{\parbox{9.0cm}{\indent \footnotesize{#1}}}
\newcommand{\tablenoteb}[1]{\parbox{17.8cm}{\indent \footnotesize{#1}}}

\newcommand{\asr}{Adv. Space Res.}

\newcommand{\astronl}{Astron. Lett.}

\newcommand{\cp}{Chem. Phys.}
\newcommand{\cpl}{Chem. Phys. Lett.}

\newcommand{\cflame}{Combust. Flame}
\newcommand{\expas}{Exp. Astron.}
\newcommand{\grs}{Geophys. Res. Lett.}

\newcommand{\ijck}{Int. J. Chem. Kinet.}

\newcommand{\jcsft}{J. Chem. Soc. Faraday Trans.}

\newcommand{\jppc}{J. Photochem. Photobiol. C}
\newcommand{\jpc}{J. Phys. Chem.}
\newcommand{\jpca}{J. Phys. Chem. A}
\newcommand{\jpcrd}{J. Phys. Chem. Ref. Data}

\newcommand{\nature}{Nature}
\newcommand{\natastro}{Nat. Astron.}
\newcommand{\pccp}{Phys. Chem. Chem. Phys.}

\newcommand{\pss}{Planet. Space Sci.}

\begin{document}

\title{Quantification of abundance uncertainties\\ in chemical models of exoplanet atmospheres}

\titlerunning{Quantification of abundance uncertainties in chemical models of exoplanet atmospheres}
\authorrunning{Ag\'undez}

\author{Marcelino~Ag\'undez}

\institute{
Instituto de F\'isica Fundamental, CSIC, Calle Serrano 123, E-28006 Madrid, Spain\\ \email{marcelino.agundez@csic.es} 
}

\date{Received; accepted}

\abstract
{Chemical models are routinely used to predict the atmospheric composition of exoplanets and compare it with the composition retrieved from observations, but little is known about the reliability of the calculated composition. We carried out a sensitivity analysis to quantify the uncertainties in the abundances calculated by a state-of-the-art chemical atmosphere model of the widely observed planets WASP-33b, HD\,209458b, HD\,189733b, WASP-39b, GJ\,436b, and GJ\,1214b. We found that the abundance uncertainties in the observable atmosphere are relatively small, below one order of magnitude and in many cases below a factor of two, where vertical mixing is a comparable or even larger source of uncertainty than (photo)chemical kinetics. In general, planets with a composition close to chemical equilibrium have smaller abundance uncertainties than planets whose composition is dominated by photochemistry. Some molecules, such as H$_2$O, CO, CO$_2$, and SiO, show low abundance uncertainties, while others such as HCN, SO$_2$, PH$_3$, and TiO have more uncertain abundances. We identified several critical albeit poorly constrained processes involving S-, P-, Si-, and Ti-bearing species whose better characterization should lead to a global improvement in the accuracy of models. Some of these key processes are the three-body association reactions S + H$_2$, Si + O, NH + N, and N$_2$H$_2$ + H; the chemical reactions S + OH $\rightarrow$ SO + H, NS + NH$_2$ $\rightarrow$ H$_2$S + N$_2$, P + PH $\rightarrow$ P$_2$ + H, and N + NH$_3$ $\rightarrow$ N$_2$H + H$_2$; and the photodissociation of molecules such as P$_2$, PH$_2$, SiS, CH, and TiO.}

\keywords{astrochemistry -- planets and satellites: atmospheres -- planets and satellites: composition -- planets and satellites: gaseous planets}

\maketitle

\section{Introduction}

The advent of telescope facilities such as JWST, Ariel, and ELT makes it possible to characterize exoplanet atmospheres to an unprecedented level of detail and put increasingly accurate constraints on the abundances of atmospheric constituents \citep{Venot2018,Tsai2023,Kempton2023,Xue2024,Fu2024,Bell2024,Sing2024,Welbanks2024,Schlawin2024a,Mukherjee2025,Palle2025}. To understand the origin of the composition retrieved and evaluate its chemical plausibility it is necessary to rely on chemical models \citep{Moses2011,Kawashima2021}, but a truthful comparison between calculated and observed abundances requires  taking into account the uncertainties in both types of quantities. While retrieval methods usually provide observed abundances with the corresponding statistical error \citep{Madhusudhan2009,Line2014}, a rigorous quantification of the uncertainties associated with the abundances calculated by chemical models has  thus far been overlooked.

Sensitivity analyses applied to chemical models have a long tradition in astrochemistry. The first attempts to estimate how uncertainties in reaction rate coefficients and photolysis rates propagate to abundances were carried out for the atmospheres of the Solar System gas giants Neptune and Saturn \citep{Dobrijevic1998,Dobrijevic2003}. Later studies focused on other astronomical environments rich in molecules, such as dark interstellar clouds \citep{Vasyunin2004,Wakelam2006,Wakelam2010a,Wakelam2010b,Byrne2024}, hot cores around protostars \citep{Wakelam2005}, protoplanetary disks \citep{Vasyunin2008}, circumstellar envelopes around evolved stars \citep{Wakelam2010b}, and the atmospheres of Titan \citep{Hebrard2006,Hebrard2007} and Triton \citep{Benne2022}. In addition to the quantification of uncertainties in the abundances, an important outcome of these sensitivity analyses is that they allow us to identify the most critical reactions, i.e., those that affect  the chemical composition most. By studying the kinetics of these reactions through experiments or theoretical calculations it is possible to improve the accuracy of the chemical model. Efforts in this sense have been carried out in the context of the atmosphere of Titan \citep{Carrasco2008,Hebrard2009,Dobrijevic2010a,Dobrijevic2010b}, dark clouds, and circumstellar envelopes \citep{Wakelam2010b}.

Here we present a sensitivity analysis applied to 1D chemical models of well-known hot and ultra-hot Jupiters and warm Neptunes. Our aim is to estimate quantitatively   the uncertainties associated with the abundances calculated by a state-of-the-art chemical model of an exoplanet atmosphere, and to identify  the most critical chemical reactions and photoprocesses in such environments so that they can be the subject of follow-up studies, and chemical models of exoplanet atmospheres can become more accurate. We are not aware of a similar sensitivity analysis applied to exoplanet atmospheres, apart from an uncertainty propagation model of GJ\,436b \citep{Venot2019}, and the preliminary study for hot Jupiters by \cite{Lira-Barria2025}.

\section{Methods} \label{sec:methods}

The sensitivity analysis was carried out using the code PACT\footnote{\tiny \url{https://github.com/marcelinoagundez/pact}. \label{foot:pact}} \citep{Agundez2025}, which calculates the 1D vertical distribution of the species in a planetary atmosphere. The atmosphere is divided into 70 layers between 100 bar and 10$^{-8}$ bar, and the initial temperature and composition (at chemical equilibrium) are solved self-consistently. The vertical resolution of 70 layers provides a good compromise between accuracy and computational cost. Afterward, thermochemical kinetics, photochemistry, and vertical mixing allow the composition to evolve with time until a steady state is reached. The temperature was not recalculated on the fly during the evolution to reduce computing time. This assumption is nevertheless justified due to the small impact on the chemical composition, as shown in \cite{Agundez2025}. The different model assumptions in the code and the data used (thermochemical data, IR opacities, chemical network, and UV cross sections) are described in \cite{Agundez2025}. Of particular relevance for this study were the chemical network and the UV cross section data, and  their associated uncertainties were of utmost importance. We were also particularly concerned with the role of vertical mixing strength, which in the case of exoplanet atmospheres is poorly constrained  \citep{Parmentier2013,Agundez2014a}.

The chemical network involves the elements H, C, N, O, S, Si, P, Ti, and He and contains 163 gaseous neutral species connected by 2351 forward chemical reactions. Reverse reactions are also included with the rate coefficients computed via detailed balance using the thermochemical data of the species involved. Thermochemical data in the form of NASA polynomials were taken mainly from the NASA/CEA database\footnote{\tiny \url{https://www1.grc.nasa.gov/research-and-engineering/ceaweb/}} \citep{McBride2002} and the Third Millenium Thermochemical Database\footnote{\tiny \url{https://burcat.technion.ac.il/}} \citep{Goos2018}. The references for the thermochemical data of each species are given in Table A.1 of \cite{Agundez2025} and the data can be publicly retrieved.$^{\ref{foot:pact}}$ The chemical network, which is publicly available$^{\ref{foot:pact}}$ \citep{Agundez2025}, was constructed from scratch by a careful inspection of the literature on chemical kinetics, either experimental studies or theoretical ones, or compilations. For simplicity, the reactions were divided into three categories, A, B, and C, according to the level of uncertainty of the rate coefficient. For those reactions studied experimentally in which the dependence on temperature is known or at least understood (type A) we assigned an uncertainty of $\Delta \log k$\,=\,0.3, that is, a factor of two in the rate coefficient $k$. Critical evaluations in the areas of combustion chemistry \citep{Baulch2005} and terrestrial atmospheric chemistry \citep{Burkholder2020} usually estimate uncertainties in the range $\Delta \log k$\,=\,0.2-0.5 over ample temperature ranges, or at best $\Delta \log k$\,=\,0.05-0.1, although this is usually restricted to temperatures around 298 K. Since the general temperature range over which the network is used to model exoplanet atmospheres is 100-4000 K, we adopted a round uncertainty value of $\Delta \log k$\,=\,0.3 for reactions of type A, which we think  is a realistic estimate. For reactions that have only been  studied theoretically (type B), we assigned an uncertainty of $\Delta \log k$\,=\,0.65 (around a factor of 4.5 in $k$), while for reactions for which the rate coefficient has been estimated or guessed based on the behavior of similar reactions or chemical intuition (type C) we adopted $\Delta \log k$\,=\,1.0, i.e., a factor of ten in the rate coefficient. The number of forward reactions in the categories A, B, and C are 800, 953, and 598, respectively. In the sensitivity analysis by \cite{Lira-Barria2025} the reaction network used is that of \cite{Venot2020}, where the estimated uncertainties in reaction rate coefficients are significantly smaller than the values adopted here, where we use the network of \cite{Agundez2025}.

The UV cross section data of the different species included consist of the wavelength-dependent photoabsorption cross section together with, in the case of molecules, the wavelength-dependent branching ratios of the different photodissociation channels. Photoabsorption cross sections were taken from the NORAD database\footnote{\tiny \url{https://norad.astronomy.osu.edu/}} \citep{Nahar2020,Nahar2024} for atoms and from the MPI-Mainz\footnote{\tiny \url{https://www.uv-vis-spectral-atlas-mainz.org/uvvis/}} \citep{Keller-Rudek2013}, Leiden\footnote{\tiny \url{https://home.strw.leidenuniv.nl/~ewine/photo/}} \citep{Heays2017,Hrodmarsson2023}, and Huebner\footnote{\tiny \url{https://phidrates.space.swri.edu/}} \citep{Huebner1992} databases. Photoionization and photodissociation yields were taken from above databases if available, otherwise we assumed a photoionization yield of 100\,\% for energies above the photoionization threshold, and a photodissociation yield of 100\,\% for energies between the photoionization and photodissociation threshold. Information on the contribution of the different photodissociation channels was obtained from the literature and from the Huebner database. For molecules with no UV data we assumed a guess value of 10$^{-18}$ cm$^{-2}$ from the photoionization threshold to 250 nm, or in the range of 100-250 nm if the ionization threshold was not known. More details are given in \cite{Agundez2025}. We proceeded in a similar way to the reaction rate coefficients and classified the data in three general categories, A, B, and C, with associated cross section uncertainties $\Delta \log \sigma$\,=\,0.3, 0.65, and 1.0, respectively. In general, category A comprises stable molecules for which the cross section has been experimentally determined, and atoms. The Leiden database \citep{Heays2017,Hrodmarsson2023} quotes uncertainties of 20-30\,\%, better than the factor of two adopted here, for the photoabsorption cross section of stable molecules, although we note that this database does not provide branching ratios for the different photodissociation channels. Species for which cross section data come from theoretical calculations are grouped into category B (usually radicals), while those for which the cross section is unknown (and thus merely guessed) are in group C. The number of species with UV cross section data in categories A, B, and C are 67, 37, and 59, respectively.

The strength of vertical mixing in the radiative part of the atmosphere (which in our case is the whole region modeled) is given by the recommendation of \cite{Moses2022}
\begin{equation}
K_{zz} = 5 \times 10^8 p^{-0.5} \bigg( \frac{H_1}{620} \bigg) \bigg( \frac{T_p}{1450} \bigg)^4, \label{eq:kzz}
\end{equation}
where $K_{zz}$ is the eddy diffusion coefficient in units of cm$^2$ s$^{-1}$, $p$ is the pressure in units of bar, $H_1$ is the atmosphere scale height at 1 mbar in units of km, and $T_p$ is the planetary equilibrium temperature in units of K, with an upper limit of 10$^{11}$ cm$^2$ s$^{-1}$. Given the dispersion of $K_{zz}$ values inferred from GCMs of hot Jupiters \citep{Parmentier2013}, we adopt an uncertainty of a factor of ten for the $K_{zz}$ profile given by Eq.\,(\ref{eq:kzz}), i.e., $\Delta \log K_{zz}$\,=\,1.0.

\begin{figure*}
\centering
\includegraphics[angle=0,width=\textwidth]{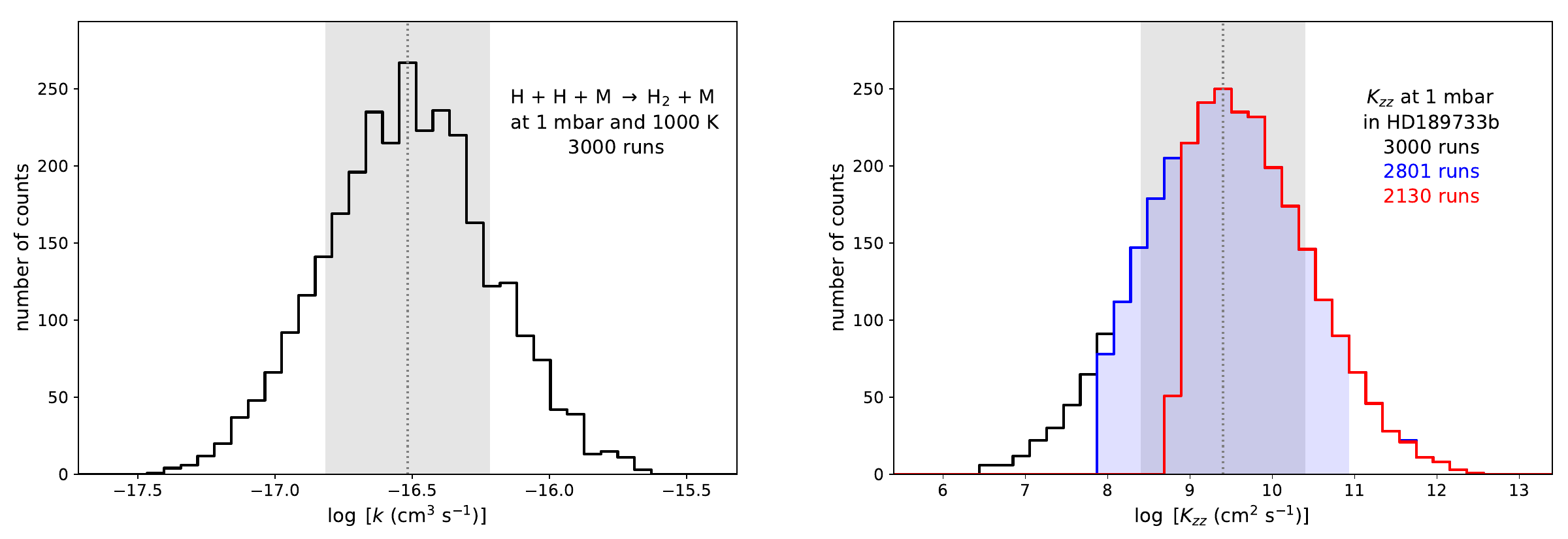}
\caption{Left panel: Random lognormal distribution of the rate coefficient of the reaction H + H + M $\rightarrow$ H$_2$ + M, evaluated at 1 mbar and 1000 K, for the models where (photo)chemical kinetics is perturbed. Right panel: Random lognormal distribution of $K_{zz}$ around the nominal value for the HD\,189733b models where vertical mixing is perturbed, where the black histogram refers to the whole set of 3000 runs, the blue one to the set of converged runs, and the red one excludes those runs with a relaxed convergence (see text). The runs actually considered for the sensitivity analysis are those converged but excluding the right wing to have a symmetric distribution (blue area). In both panels, the vertical dotted line indicates the nominal value and the shaded gray area the range $\pm$\,1\,$\sigma$ around it.} \label{fig:histogram}
\end{figure*}

Once uncertainties in the variable parameters (reaction rate coefficients $k$, UV cross sections $\sigma$, and eddy diffusion coefficient $K_{zz}$) are specified, the rationale behind the sensitivity analysis consists of running a high number of 1D models (typically $>$\,1000) in which the variable parameters $k$, $\sigma$, and $K_{zz}$ are randomly varied within their uncertainties using a lognormal distribution (e.g., \citealt{Dobrijevic1998,Wakelam2005}; see Fig.\,\ref{fig:histogram}). In order to isolate the uncertainties associated with (photo)chemistry and vertical mixing separately, we ran two independent series of 3000 runs. We  verified that the results are not sensitive to the number of runs once these are above 1000. In the first series, only reaction rate coefficients and UV cross sections were randomly varied and $K_{zz}$ was kept fixed, and in the second series the (photo)chemical parameters were kept fixed to their nominal values and $K_{zz}$ was randomly varied. The standard deviation (1\,$\sigma$) is given by $\Delta \log k$, $\Delta \log \sigma$, or $\Delta \log K_{zz}$, and random perturbation factors larger than 3\,$\sigma$ were neglected. We adopted the same random perturbation for forward and reverse reaction (to ensure a detailed balance) and for all fragmentation channels in the photodissociation of a given species. Most runs reach the steady state without problems, but some have difficulties using the convergence criterion adopted in \cite{Agundez2025}. To avoid exceedingly large computing times for problematic runs, we considered that steady state is also reached if the integration time is longer than the steady state time of the unperturbed run and the relative abundance variation has fallen below one. Even   with this more relaxed convergence criterion, some models did not reach steady state due to numerical issues, and were thus neglected. The fraction of successful runs varied from one case to another, but it was always above 75\,\% and often higher than 95\,\%. In the series of models in which only $K_{zz}$ was randomly varied, nonconverged runs tended to happen preferentially at one side, either at lower $K_{zz}$ or at higher $K_{zz}$ depending on the planet. This caused a nonsymmetric distribution in which the wing of one of the sides of the lognormal distribution which should be symmetric (see Fig.\,\ref{fig:histogram}) is truncated. To avoid this behavior and to retain a symmetric distribution we skipped those runs in which the absolute value of the variation is above a given threshold. After all models were run, it was straightforward to compute the mean and the standard deviation of the steady state abundance of every species at each altitude. If the runs converged under the relaxed convergence criterion are excluded, the statistics are reduced but the abundance uncertainties due to (photo)chemical kinetics are only marginally affected. However, excluding these runs affects the uncertainties due to vertical mixing because they occur almost exclusively on one side of the lognormal distribution (compare the blue and red histograms in the right panel of Fig.\,\ref{fig:histogram}),  and thus excluding them results in a markedly asymmetrical distribution of $K_{zz}$, even if the wings are truncated. Therefore, calculated uncertainties are more reliable when including runs converged under relaxed conditions.

An important outcome of the sensitivity analysis is the identification of the most critical (photo)reactions. To that purpose, we  evaluated the correlation between the perturbations in the (photo)reaction rate coefficients and the variations induced in the abundances of the different species. We define the quantities $X_i^l$\,=\,$\log k_i^l - \log k_i^0$, where $k_i ^l$ and $k_i^0$ stand for the rate coefficient of the (photo)reaction $i$ (perturbed in run $l$ and unperturbed, respectively), and $Y_{j,k}^l$\,=\,$\log f_{j,k}^l - \overline{\log f_{j,k}}$, where $f_{j,k}^l$ and $\overline{f_{j,k}}$ are the mole fraction of species $j$ at layer $k$ (resulting from run $l$ and the mean value among all runs, respectively). To quantify the correlation between a (photo)reaction $i$ and a species $j$ we use the Pearson and Spearman correlation coefficients. The Pearson correlation coefficient is evaluated as \citep{Wakelam2010b}
\begin{equation}
P_{j,k}^i = \frac{\sum_{l=1}^n X_i^l \, Y_{j,k}^l}{\sqrt{\sum_{l=1}^n \big(X_i^l\big)^2 \, \sum_{l=1}^n \big(Y_{j,k}^l\big)^2}},
\end{equation}
where the sums extend over all $n$ runs. The Spearman correlation coefficient is defined as
\begin{equation}
S_{j,k}^i = \frac{\sum_{l=1}^n R[X]_i^l \, R[Y]_{j,k}^l}{\sqrt{\sum_{l=1}^n \big(R[X]_i^l\big)^2 \, \sum_{l=1}^n \big(R[Y]_{j,k}^l\big)^2}},
\end{equation}
where $R[X]_i^l$ and $R[Y]_{j,k}^l$ are the ranks of the variables $X_i^l$ and $Y_{j,k}^l$, and here it is evaluated using the routine \texttt{spear} from Fortran Numerical Recipes \citep{Press1992}. To avoid outliers with minimal abundance variations we set $P_{j,k}^i$ and $S_{j,k}^i$ to zero if the average of $Y_{j,k}^l$ over all runs is smaller than 0.01. Since we are mainly interested in the impact of (photo)reactions on abundant species, we skip those cases where $f_{j,k}^l$ is below 10$^{-10}$ and set the corresponding correlation coefficient $P_{j,k}^i$ and $S_{j,k}^i$ to zero. We rank (photo)reactions in order of importance according to their global correlation coefficient, which is calculated as
\begin{equation} \label{eq:p}
C^i = \sum_j |C_j^i|,
\end{equation}
where $|C_j^i|$ stands for the maximum value of $|P_{j,k}^i|$ or $|S_{j,k}^i|$ among all layers $k$, depending on whether the Pearson or Spearman coefficient is used, and the sum is restricted to |$C_j^i$|\,$>$\,0.3.

\section{Planets} \label{sec:planets}

\begin{table}
\small
\caption{Star and planet parameters.}
\label{table:planets}
\centering
\begin{tabular}{l@{\hspace{0.38cm}}c@{\hspace{0.38cm}}c@{\hspace{0.38cm}}c@{\hspace{0.38cm}}c@{\hspace{0.38cm}}c@{\hspace{0.38cm}}c}
\hline \hline
Planet             & $R_\star$ & $T_\star$      & $a$      & $M_p$ & $R_p$ & $T_p$\\
                       & (R$_\odot$) & (K)      & (mAU)      & (M$_\oplus$) & (R$_\oplus$) & (K) \\
\hline
WASP-33b\,$^a$    & 1.444 & 7400 & 25.65 & 667.4        & 17.97 & 2677 \\
HD\,209458b\,$^b$ & 1.203 & 6092 & 47.47 & 226.9        & 15.468 & 1479 \\
HD\,189733b\,$^c$ & 0.805 & 4875 & 31   & 371.9        & 12.902 & 1198 \\
WASP-39b\,$^d$    & 0.895 & 5400 & 48.6 & 89.0        & 14.2 & 1117 \\
GJ\,436b\,$^e$    & 0.464 & 3684 & 28.87 & 18.55        & 3.645 & 712 \\
GJ\,1214b\,$^f$   & 0.216 & 3026 & 14.11 & 8.17         & 2.742 & 571 \\
\hline
\end{tabular}
\tablenotea{\\
Note.-- $R_\star$ and $T_\star$ are the stellar radius and effective temperature, respectively, $a$ is the planet-star distance, and $M_p$, $R_p$, and $T_p$ are the planetary mass, radius, and equilibrium temperature, respectively. The equilibrium temperature is computed as $T_p$\,=\,$\sqrt{R_\star/(2 a)}$ $T_\star$, where the albedo and emissivity are implicitly assumed to be 0 and 1, respectively.\\
$^a$\,\cite{CollierCameron2010}, \cite{Smith2011}, \cite{Lehmann2015}, \cite{Hardy2015}. $^b$\,\cite{Boyajian2015}, \cite{Southworth2010}. $^c$\,\cite{Boyajian2015}, \cite{Paredes2021}. $^d$\,\cite{Faedi2011}. $^e$\,\cite{Torres2007}, \cite{Southworth2010}, \cite{Melo2024}. $^f$\,\cite{Harpsoe2013}, \cite{Cloutier2021}.
}
\end{table}

We performed the sensitivity analysis for some of the most widely observed exoplanet atmospheres, which fall into the category of highly irradiated gas giants. Concretely, we selected the ultra-hot Jupiter WASP\,33b, which has an equilibrium temperature in excess of 2500 K; the hot Jupiters HD\,209458b, HD\,189733b, and WASP-39b, with equilibrium temperatures in the range 1000-2000 K; and the warm Neptunes GJ\,436b and GJ\,1214b, whose equilibrium temperatures are in the range 500-1000 K.

\begin{figure}
\centering
\includegraphics[angle=0,width=\columnwidth]{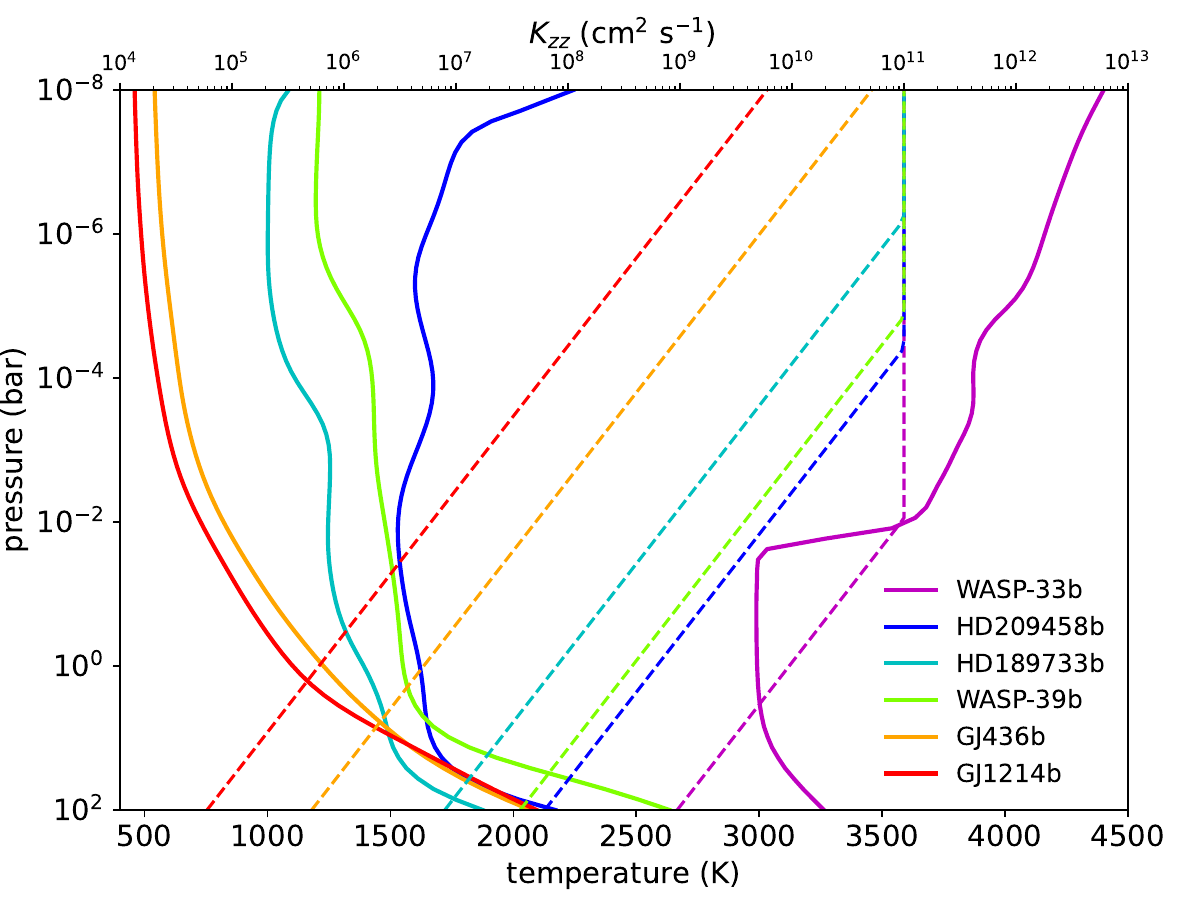}
\caption{Calculated pressure-temperature profiles (solid lines referred to the bottom $x$-axis) and adopted $K_{zz}$ vertical profiles (dashed lines referred to the top $x$-axis) for the six exoplanets investigated.} \label{fig:temp_kzz}
\end{figure}

The parameters adopted to model the dayside atmospheres of these six exoplanets are given in Table\,\ref{table:planets}, where they are ordered according to their equilibrium temperatures. The temperature and $K_{zz}$ vertical profiles of the six planets are shown in Fig.\,\ref{fig:temp_kzz}. For all planets, the Bond albedo and emissivity of the bottom boundary were assumed to be 0 and 1, respectively, the zenith stellar illumination angle is 48$^\circ$ for the computation of temperature and photochemistry \citep{Tsai2021}, the internal temperature was assumed to be 300 K \citep{Ginzburg2015,Thorngren2019,Komacek2022}, and the heat was assumed to be efficiently redistributed from the dayside to the nightside, except for WASP-33b, in which case we assumed no dayside-to-nightside heat redistribution \citep{Zhang2018,Koll2022,Dang2025}. The metallicity was assumed to be solar for HD\,209458b and HD\,189733b \citep{Xue2024,Finnerty2024}, 10\,$\times$\,solar for WASP-33b and WASP-39b \citep{Cont2022,Tsai2023}, 100\,$\times$\,solar for GJ\,436b \citep{Mukherjee2025}, and 1000\,$\times$\,solar for GJ\,1214b \citep{Schlawin2024a}. For the warm Neptunes GJ\,436b and GJ\,1214b we did not include the elements Si and Ti because their refractory character makes it likely that they are condensed out of the gas phase. The stellar spectra were built from different sources (see details in \citealt{Agundez2025}). For the UV-visible part, we used the MUSCLES\footnote{\tiny \url{https://archive.stsci.edu/prepds/muscles/} \label{muscles}} database \citep{France2016,Youngblood2016,Loyd2016,Behr2023}, where spectra for GJ\,436 and GJ\,1214 are available, and for WASP-33 and WASP-39 the spectra of WASP-17 and WASP-77A, respectively, were adopted as proxies. In the cases of HD\,209458 and HD\,189733 we used that of the Sun (WHI; \citealt{Woods2009}) and that of $\epsilon$\,Eridani, respectively, as proxies based on their similar stellar properties. For the infrared part, we used the Kurucz\footnote{\tiny \url{http://kurucz.harvard.edu/stars.html} \label{kurucz}} and Castelli-Kurucz\footnote{\tiny \url{https://wwwuser.oats.inaf.it/fiorella.castelli/grids.html} \label{castelli-kurucz}} databases \citep{Castelli2004}, while at wavelengths longer than 160-300 $\mu$m we assumed a blackbody spectral shape.

\section{Results and discussion} \label{sec:results}

\subsection{Abundance uncertainties} \label{subsec:abundance_uncertainties}

\begin{figure*}
\centering
\includegraphics[angle=0,width=\textwidth]{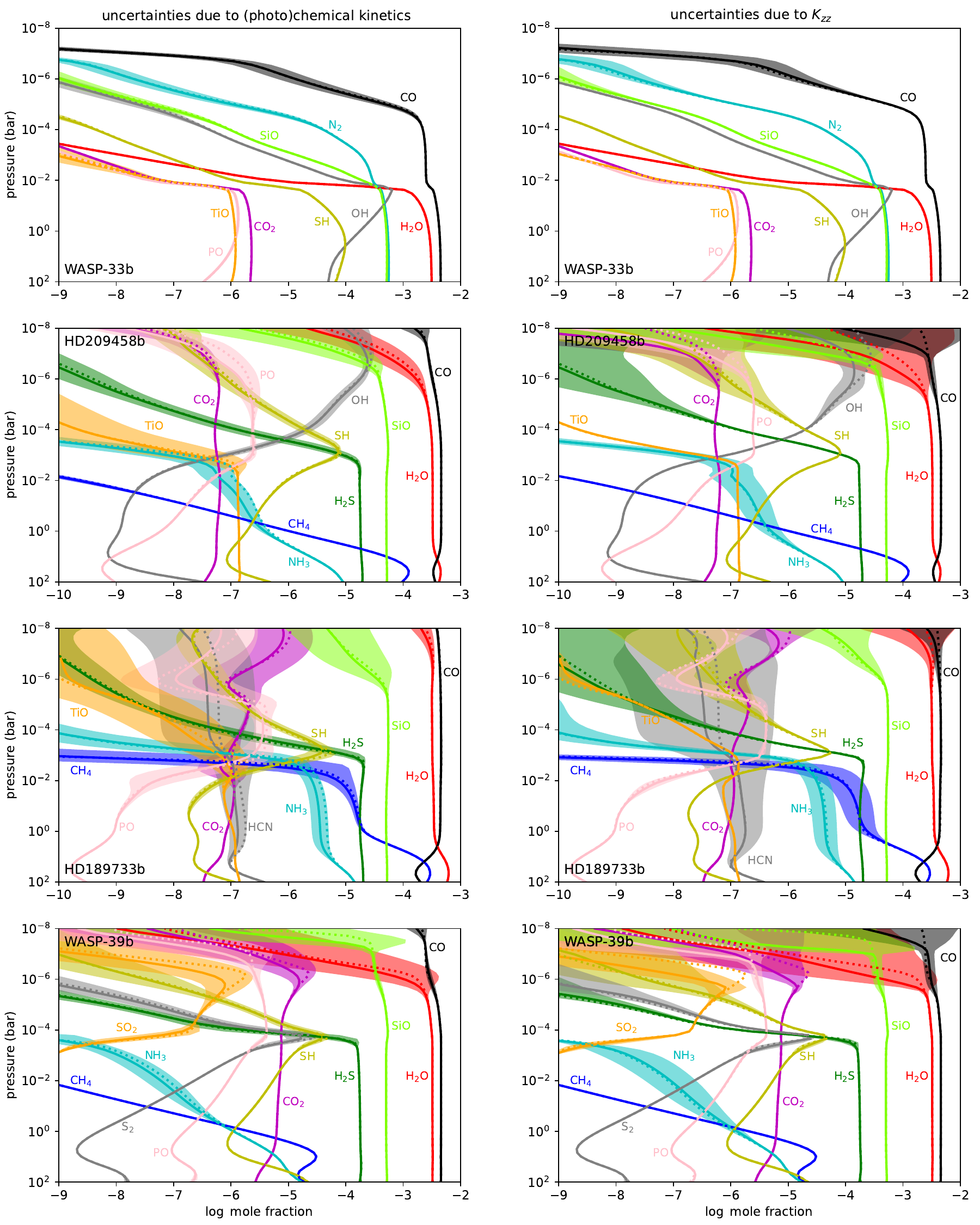}
\caption{Calculated vertical distribution of abundances in the ultra-hot Jupiter WASP-33b and the hot Jupiters HD\,209458b, HD\,189733b, and WASP-39b. The solid lines correspond to the mean abundance, the shaded areas to the range around the mean $\pm$\,$\sigma$, and the dotted lines to the abundances resulting from the unperturbed model. The panels on the left show the effect on the calculated abundances of the uncertainties on (photo)chemical kinetics, while the right panels correspond to the effect on the abundances of the uncertainty in $K_{zz}$.} \label{fig:hot_jupiters}
\end{figure*}

\begin{figure*}
\centering
\includegraphics[angle=0,width=\textwidth]{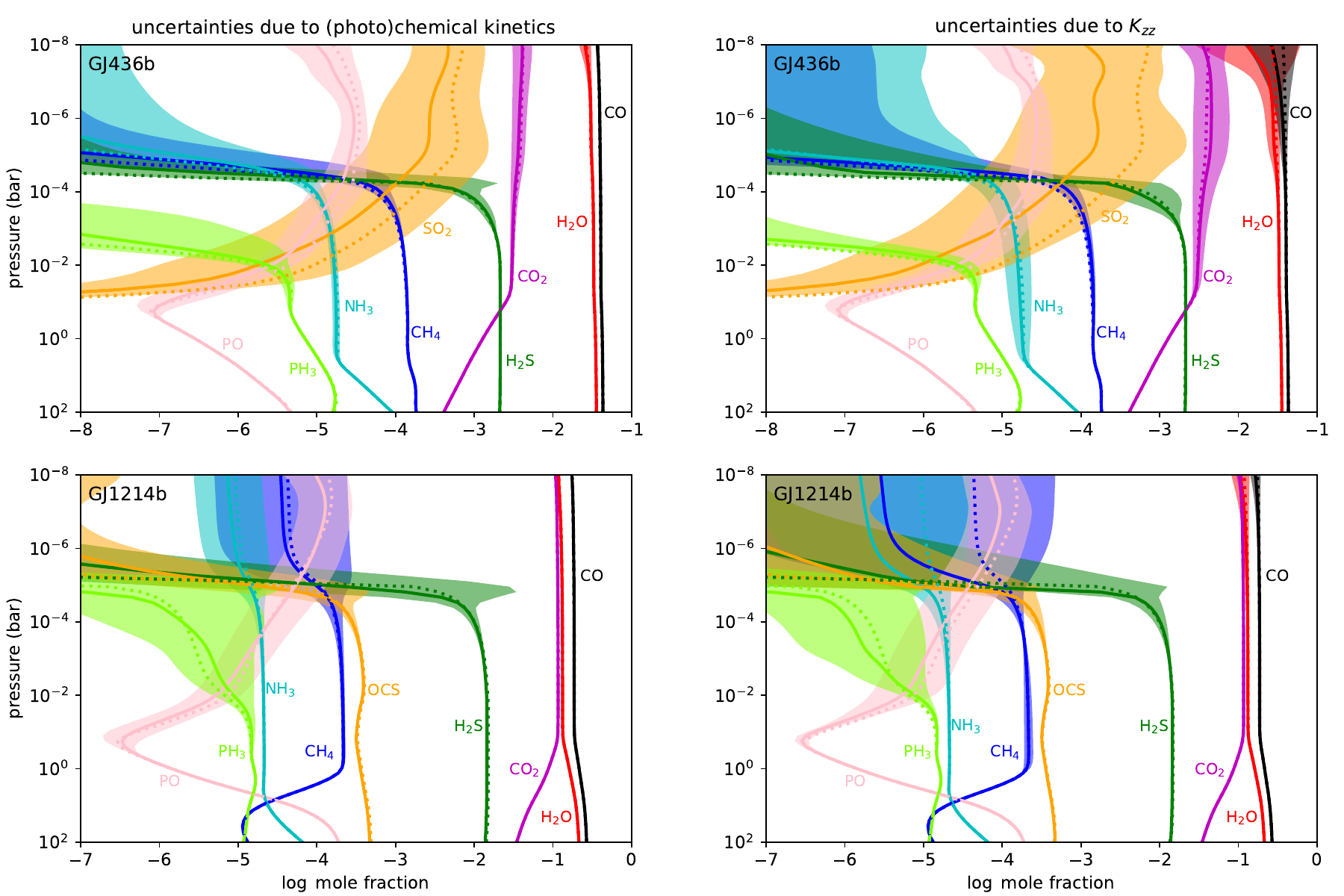}
\caption{Same as in Fig.\,\ref{fig:hot_jupiters}, but for the warm Neptunes GJ\,436b and GJ\,1214b.} \label{fig:warm_neptunes}
\end{figure*}

The results from the sensitivity analysis are shown in Fig.\,\ref{fig:hot_jupiters} for the ultra-hot Jupiter WASP\,33b and the hot Jupiters HD\,209458b, HD\,189733b, and WASP-39b, and in Fig.\,\ref{fig:warm_neptunes} for the warm Neptunes GJ\,436b and GJ\,1214b. The left panels refer to the uncertainties caused by chemical kinetics (reaction rate coefficients and UV cross sections), while the right panels correspond to the uncertainties introduced by the strength of vertical mixing ($K_{zz}$). Other sources of uncertainty that could affect the abundances calculated by a chemical model are those associated with thermochemical data, incompleteness of the chemical network, presence of condensates, departure from 1D geometry, elemental composition, stellar irradiation, or atmospheric temperature, although none of them are taken into account here. In all cases in Fig.\,\ref{fig:hot_jupiters} and Fig.\,\ref{fig:warm_neptunes} the solid lines correspond to the mean abundance profiles, the shaded areas to the range of abundances around the mean $\pm$\,$\sigma$, and the dotted lines to the abundance profiles in the unperturbed case.

In general, the calculated abundances in the lower part of the atmosphere (below the 1 bar pressure level) are close to chemical equilibrium and thus are not affected by uncertainties in (photo)chemical kinetics or $K_{zz}$. Only uncertainties in the thermochemical data would affect them, but they are not taken into account here. This finding applies to all planets investigated, independently of how warm or cool the atmosphere is. The observational relevance is  limited, however, because these regions are usually not probed by emission or transmission spectroscopic observations. We are thus mainly interested in the abundance uncertainties of species present in the observable atmosphere, which typically lies in the 1-10$^{-6}$ bar pressure range. In this region, abundance uncertainties remain relatively small, less than one order of magnitude and for many species lower than a factor of two. In general terms we can state that in those planets whose overall atmospheric composition is close to chemical equilibrium the calculated abundances are little affected by uncertainties in (photo)chemical kinetics or vertical mixing. On the other hand, in those planets where photochemistry plays an important role in establishing the atmospheric composition, the calculated abundances have sizable uncertainties. We find that this dichotomy between composition driven by either chemical equilibrium or photochemistry, rather than other factors such as the equilibrium temperature of the planet, is the main element that determines the level of uncertainty in the calculated composition. The ultra-hot Jupiter WASP-33b illustrates very well  that when the composition is close to chemical equilibrium \citep{Agundez2025}, the calculated abundances have very small uncertainties, while HD\,189733b shows that when the photochemistry is important, the abundances become more uncertain (see Fig.\,\ref{fig:hot_jupiters}). A general finding  is that uncertainties due to vertical mixing are comparable to, and in many cases larger than, those associated with (photo)chemical kinetics. Therefore, reducing the uncertainty in $K_{zz}$ is as important as improving the accuracy of reaction rate coefficients and photo cross sections. It is noteworthy that the mean abundance does not always coincide with that from the unperturbed model, with the most significant differences being found in the upper atmosphere where photochemistry dominates. This behavior reflects the fact that the impact of varying (photo)chemical or eddy coefficients on the abundances is nonlinear and nonsymmetric. For example, an increase in a given coefficient may influence to a large extent the abundance of a given species, but a decrease in that coefficient may leave that abundance almost unchanged. That is, although the input distribution in the variation of rate coefficients or eddy diffusion coefficients is lognormal and thus symmetrical, the resulting dispersion of abundances often departs from a symmetrical lognormal distribution, and thus the mean abundance does not necessarily coincide with the unperturbed one.

There are also some general trends depending on the species. Within a given planetary atmosphere, not all species have similar abundance uncertainties. In general, those species that tend to remain uniform with altitude show little abundance dispersion. This applies to CO, H$_2$O, N$_2$, and SiO, molecules that are very stable and act as major reservoirs of carbon, oxygen, nitrogen, and silicon, respectively. Only in the more tenuous upper layers (above the 10$^{-6}$ bar pressure level), where these species start to be destroyed by photodissociation, there may be an increase in their abundance uncertainty, although these layers are usually out of the range probed by observations. That is, molecules such as H$_2$O and CO, which are routinely observed in hot and warm gas giant exoplanets \citep{Sing2016,Yan2022,Yang2024,Mraz2024,Blain2024a,Xue2024,Finnerty2024,Inglis2024,Beatty2024,Zhang2025}, are not much affected by uncertainties in (photo)reaction networks or vertical mixing, and thus can be robustly used to constrain parameters related to the elemental composition such as the C/O ratio \citep{Madhusudhan2012,Moses2013a,Molliere2015}. It is also interesting to note that if silicon is not depleted from the gas phase due to condensation as quartz or silicates, silicon monoxide (SiO) is predicted to be the major carrier of silicon and its predicted abundance is rather accurate and free of uncertainties from (photo)chemistry or vertical mixing. The recent evidence of SiO in the ultra-hot Jupiters WASP-187b and WASP-121b \citep{Lothringer2022,Gapp2025,Evans-Soma2025} suggests that rather high temperatures, above 2000 K, are necessary to maintain silicon in the gas phase.

There are other species that tend to have rather uncertain abundances, regardless of the planet. In this category we find molecules such as HCN, SO$_2$, PH$_3$, and TiO. All of them have in common that their abundances at pressures below 1 mbar are largely controlled by photochemistry. In the case of SO$_2$ this is because it is essentially formed by photochemistry and in the cases of HCN, PH$_3$, and TiO it is because of their destruction by UV photons. The ultimate cause of the uncertainty associated with the abundance of these molecules is that the chemical kinetics and/or photo cross sections related to them are poorly known. This means that their abundances should be more accurately predicted once the corresponding (photo)chemical data are better known.

Apart from the general trends found for some molecules and discussed above, there are also species present in the observable atmosphere whose abundance uncertainty depends largely on the specific conditions of each planet. For example, CO$_2$ has small abundance uncertainties below the 10$^{-6}$ bar pressure level in most exoplanets. Only in HD\,189733b and GJ\,436b do we find that abundance errors can be up to a factor of a few. The low abundance uncertainty found for CO$_2$ is interesting because it validates the use of this molecule as an indicator of the metallicity, as has been previously recognized \citep{Zahnle2009,Line2011,Moses2013b,Agundez2014b,Venot2014} and it is currently routinely applied \citep{JWST2023,Xue2024,Beatty2024,Schlawin2024a,Schlawin2024b,Ohno2025,Mukherjee2025,Balmer2025}. Methane (CH$_4$) and hydrogen sulfide (H$_2$S) are usually present below the 10$^{-5}$-10$^{-4}$ bar pressure level, where they have a very low abundance uncertainty -- the largest is just a factor of two for CH$_4$ in HD\,189733b. Ammonia (NH$_3$) has abundance uncertainties ranging from a factor of two to one order of magnitude depending on the planet, while phosphorus monoxide (PO)  also has a mild planet-dependent behavior, with rather accurate abundances in WASP-33b, HD\,209458b, and WASP-39b, and abundance uncertainties of up to one order of magnitude in HD\,189733b, GJ\,436b, and GJ\,1214b.

Ideally, the goodness of an exoplanet atmosphere chemical model should rely on the comparison between the abundances calculated and those retrieved from observations, taking into account their corresponding uncertainties. However, in many cases observational studies do not retrieve the abundances themselves, but parameters such as the metallicity and/or the C/O elemental ratio \citep{Cont2022,JWST2023,Xue2024,Zhang2025}, making difficult to attempt the aforementioned comparison exercise. In the future it would be desirable to test the reliability of state-of-the-art chemical models on statistical grounds by making a rigorous comparison between calculated abundances at the pressure level probed by observations and the values retrieved from observations, with their corresponding uncertainties. However, it is still relatively frequent that different observational studies end up with conflicting results regarding the detection of a molecule itself, and therefore for the moment we have to discuss our results in the context of the state-of-the-art observational results. We   list in Table\,\ref{table:abundances} the molecular abundances retrieved from observations for the planets under study here. The compilation is non-exhaustive because it focuses on molecules confidently detected from the most recent observational studies. No molecules are reported in GJ\,436b and GJ\,1214b because there is little evidence for them from recent JWST data (see below).

\begin{table*}
\footnotesize
\caption{Observed molecular abundances.}
\label{table:abundances}
\centering
\begin{tabular}{lllcl}
\hline \hline
Planet             & Molecule & Region & Abundance\,$^a$ & Comment \\
\hline
WASP-33b    & H$_2$O & Dayside & [$-$5.0, $-$4.0] & Spectrophotometry HST \citep{Haynes2015}. \\
                      & & Limb & & High spectral resolution, no abundance retrieved \citep{Yang2024}. \\
\cline{2-5}
                     & CO & Nightside & [$-$5.0, $-$1.6] & High spectral resolution \citep{Mraz2024}. \\
                     & & Dayside & & High spectral resolution, no abundance retrieved \citep{Yan2022}. \\
                     & & Dayside & & High spectral resolution, no abundance retrieved \citep{vanSluijs2023}. \\
                     & OH & Dayside & & High spectral resolution, no abundance retrieved \citep{Nugroho2021}. \\
                     & TiO & Dayside & & High spectral resolution, no abundance retrieved \citep{Cont2021}. \\
\hline
HD\,209458b & H$_2$O & Dayside & [$-$5.0, $-$2.0] & Spectrophotometry HST and Spitzer \citep{Line2016}. \\
                      & & Limb & [$-$5.0, $-$4.3] & Spectrophotometry HST and Spitzer \citep{Pinhas2019}. \\
                      & & Limb & & High spectral resolution, no abundance retrieved \citep{Blain2024a}. \\
                      & & Limb & & Spectrophotometry JWST, no abundance retrieved \citep{Xue2024}. \\
\cline{2-5}
                      & CO$_2$ & Limb & & Spectrophotometry JWST, no abundance retrieved \citep{Xue2024}. \\
\hline
HD\,189733b & H$_2$O & Dayside & [$-$5.3, $-$4.1] & Spectrophotometry JWST \citep{Inglis2024}. \\
                      & & Dayside & [$-$2.8, $-$2.5] & Spectrophotometry JWST \citep{Zhang2025}. \\
                      & & Dayside & [$-$3.4, $-$2.6] & High spectral resolution \citep{Finnerty2024}. \\
                      & & Limb & [$-$5.3, $-$4.6] & Spectrophotometry HST and Spitzer \citep{Pinhas2019}. \\
                      & & Limb & [$-$4.2, $-$2.9] & Spectrophotometry JWST \citep{Fu2024}. \\
                      & & Limb & [$-$3.2, $-$2.8] & Spectrophotometry JWST \citep{Zhang2025}. \\
                      & & Limb & [$-$2.6, $-$2.3] & High spectral resolution \citep{Blain2024b}. \\
                      & & Limb & [$-$4.4, $-$3.2] & High spectral resolution \citep{Klein2024}. \\
                      & & Limb & [$-$4.8, $-$4.0] & High spectral resolution \citep{Boucher2021}. \\
\cline{2-5}
                      & CO & Dayside & [$-$2.8, $-$2.3] & Spectrophotometry JWST \citep{Zhang2025}. \\
                      & & Dayside & [$-$3.8, $-$2.8] & High spectral resolution \citep{Finnerty2024}. \\
                      & & Limb & [$-$4.7, $-$3.3] & Spectrophotometry JWST \citep{Fu2024}. \\
                      & & Limb & [$-$3.2, $-$2.8] & Spectrophotometry JWST \citep{Zhang2025}. \\
\cline{2-5}
                      & CO$_2$ & Limb & [$-$6.7, $-$5.6] & Spectrophotometry JWST \citep{Fu2024}. \\
\cline{2-5}
                      & H$_2$S & Dayside & [$-$4.1, $-$3.1] & Spectrophotometry JWST \citep{Inglis2024}. \\
                      & & Dayside & [$-$8.4, $-$5.1] & High spectral resolution \citep{Finnerty2024}. \\
                      & & Limb & [$-$4.9, $-$3.9] & Spectrophotometry JWST \citep{Fu2024}. \\
\hline
WASP-39b    & H$_2$O &Limb &  [$-$4.9, $-$3.4] & Spectrophotometry HST and Spitzer \citep{Pinhas2019}. \\
                      & & Limb & & Spectrophotometry JWST, no abundance retrieved \citep{Rustamkulov2023}. \\
\cline{2-5}
                      & CO & Limb & & Spectrophotometry JWST, no abundance retrieved \citep{Rustamkulov2023}. \\
\cline{2-5}
                      & CO$_2$ & Limb & & Spectrophotometry JWST, no abundance retrieved \citep{Rustamkulov2023}. \\
\cline{2-5}
                     & SO$_2$ & Limb & [$-$6.3, $-$4.6] & Spectrophotometry JWST \citep{Powell2024}. \\
\hline
\end{tabular}
\tablenoteb{\\
Note.-- $^a$\,Abundance expressed as the decimal logarithm of the mole fraction.
}
\end{table*}

WASP-33b. This ultra-hot Jupiter stands out among the six planets investigated here because of its extremely high dayside temperatures, in excess of 3000 K (see Fig.\,\ref{fig:temp_kzz}). The composition is close to chemical equilibrium \citep{Agundez2025} and simple molecules are abundant in the lower atmosphere, while atoms dominate in the upper layers. This picture is in good agreement with the observations. Molecules such as H$_2$O and CO have been detected on the cooler nightside \citep{Yang2024,Mraz2024}, while on the dayside metal atoms such as Ti, V, Fe, and Si and oxides such as TiO, OH, and CO, which are particularly resilient to high temperatures, have been detected \citep{Cont2021,Cont2022,Nugroho2021,Yan2022}. The abundance of H$_2$O is constrained to the range 10$^{-5}$-10$^{-4}$ from eclipse HST observations \citep{Haynes2015}, which is consistent with the sharp decline of abundance predicted in the 1-10$^{-3}$ bar region probed by these observations. On the other hand, the abundance of CO on the nightside is poorly constrained, between 10$^{-5}$ and 2.5\,$\times$\,10$^{-2}$ \citep{Mraz2024}. The fact that in ultra-hot Jupiters the calculated abundances have low uncertainties, and the likely absence of condensates due to the very high temperatures, means that chemical models can be reliably used to constrain parameters such as metallicities and relative elemental ratios.

The hot Jupiters HD\,209458b and HD\,189733b have similar metallicities (around solar) and temperatures (see Fig.\,\ref{fig:temp_kzz}), although the composition of HD\,209458b is somewhat closer to chemical equilibrium than that of HD\,189733b \citep{Venot2012,Agundez2025}. This means that in general the abundance uncertainties are smaller in HD\,209458b than in HD\,189733b, although as discussed previously, the exact behavior depends on each particular species.

HD\,209458b. Recent ground and JWST observations point to the presence of H$_2$O and CO$_2$ \citep{Xue2024,Blain2024a}, although no abundance was retrieved. CO has been also reported in HD\,209458b \citep{Hawker2018,Giacobbe2021}, although its presence is more uncertain because there is no evidence in the more recent observational studies by \cite{Xue2024} and \cite{Blain2024a}. The abundances of H$_2$O and CO$_2$ calculated with the chemical model at the pressures probed by observations have very low uncertainties, significantly smaller than those provided by observations (see Table\,\ref{table:abundances}). For example, the observed abundance of H$_2$O on the dayside is in the range 10$^{-5}$-10$^{-2}$ \citep{Line2016}. Other molecules such as TiO, CH$_4$, C$_2$H$_2$, HCN, and NH$_3$ have been also reported in HD\,209458b \citep{Desert2008,Giacobbe2021}, although these detections were later   disputed \citep{Casasayas-Barris2021,Xue2024}.

HD\,189733b. Several molecules, such as H$_2$O, CO, CO$_2$, and H$_2$S, have been reported in this hot Jupiter \citep{Inglis2024,Finnerty2024,Fu2024,Zhang2025}. Retrieved abundances for H$_2$O and CO vary depending on the type of observational data (transit or eclipse, ground high spectral resolution or space spectrophotometry) and retrieval method used. The observed abundance of H$_2$O ranges from 5\,$\times$\,10$^{-6}$ to 5\,$\times$\,10$^{-3}$, while that of CO spans from 2\,$\times$\,10$^{-5}$ to 5\,$\times$\,10$^{-3}$ (see Table\,\ref{table:abundances}). The calculated abundances of H$_2$O and CO are rather uniform with altitude and quite accurate (see Fig.\,\ref{fig:hot_jupiters}), and are fully consistent with the ranges of values derived from observations. For CO$_2$, \cite{Fu2024} derive an abundance between 2\,$\times$\,10$^{-7}$ and 3\,$\times$\,10$^{-6}$, somewhat higher than the value predicted by the chemical model, around 10$^{-7}$ (see Fig.\ref{fig:hot_jupiters}). The slight disagreement between the calculated and observed abundances is unlikely to arise from uncertainties in chemical network or eddy diffusion coefficient because the calculated abundance of CO$_2$ has a low uncertainty below the 10$^{-5}$ bar pressure level. It may reflect a higher metallicity compared to the solar value adopted in the model. JWST observations indicate that CH$_4$ is depleted in HD\,189733b compared to the prediction of chemical equilibrium \citep{Fu2024,Zhang2025}. The marked depletion predicted for CH$_4$ above the 10$^{-2}$ bar level due to photodissociation (see Fig.\,\ref{fig:hot_jupiters}) may be the cause of the low inferred abundance of methane. H$_2$S is observed with a rather high abundance in HD\,189733b, in the range 10$^{-5}$-10$^{-3}$ according to JWST observations \citep{Fu2024,Inglis2024}. Our calculations indeed predict that H$_2$S locks most of the sulfur below the 10$^{-3}$ bar pressure level, with a rather small uncertainty, in agreement with observations.

WASP-39b. This hot Jupiter provides a nice environment where  chemical models of exoplanet atmospheres can be tested, due to the availability of several interesting molecules detected, such as H$_2$O, CO$_2$, CO, and SO$_2$ \citep{JWST2023,Rustamkulov2023,Alderson2023,Tsai2023,Powell2024,Carter2024}. The observations of H$_2$O, CO$_2$, and CO are consistent with these molecules being present with their chemical equilibrium abundances, although in the aforementioned studies the retrieved parameters are usually the metallicity and the C/O ratio, rather than molecular abundances, which makes it difficult to attempt a direct comparison between the chemical model and the observations. Only in the case of H$_2$O has the abundance   been constrained to be between 10$^{-5}$ and 4\,$\times$\,10$^{-4}$ \citep{Pinhas2019}, which is lower than the abundance predicted by the chemical model, around 3\,$\times$\,10$^{-3}$ (see Fig.\ref{fig:hot_jupiters}). The uncertainty in the calculated abundance of H$_2$O is negligible below the 10$^{-5}$ bar pressure level and therefore this cannot be the cause of the disagreement. In the case of SO$_2$, the chemical equilibrium abundance is far below the amount needed to account for the observations, which are well explained by a photochemical origin of SO$_2$ \citep{Tsai2023}. Indeed, SO$_2$ is enhanced in the upper atmosphere, between 10$^{-4}$ and 10$^{-6}$ bar with abundances between 10$^{-7}$ and 3\,$\times$\,10$^{-6}$ (see Fig.\,\ref{fig:hot_jupiters}). It is interesting to note that the uncertainty in the calculated abundance of SO$_2$ is around a factor of five, which is somewhat larger than the dispersion found from the comparison between different photochemical models presented in \cite{Tsai2023}. Our calculations indicate that there is margin for a better understanding of the (photo)chemistry of SO$_2$ and a reduction of the uncertainty in its calculated abundance (see Sect.\,\ref{subsec:critical_reactions}).

GJ\,436b and GJ\,1214b. These warm Neptunes are different from the hot Jupiters discussed above in that extremely high metallicities and clouds seem to be the rule. Previous Spitzer observations of the dayside emission spectrum of GJ\,436b indicated the presence of H$_2$O, CH$_4$, CO, and CO$_2$ \citep{Stevenson2010,Madhusudhan2011,Line2014}. However, more recent JWST observations show no conclusive evidence of any particular molecule in the emission spectrum, except for  weak evidence at 2$\sigma$ of CO$_2$ \citep{Mukherjee2025}. These observations, together with previous HST observations of the transmission spectrum, which appears rather featureless \citep{Knutson2014,Lothringer2018}, point to an atmosphere with a high metallicity (in the range 100-1000\,$\times$\,solar) and possibly optically thick clouds or hazes. Our calculations indicate that for metallicities of $\sim$100\,$\times$\,solar, as adopted for GJ\,436b, the uncertainties in the abundances due to either (photo)chemical kinetics or vertical mixing are small for most observable molecules, such as H$_2$O, CO, CO$_2$, CH$_4$, and H$_2$S (see Fig.\,\ref{fig:warm_neptunes}). The chemical model predicts that SO$_2$ could reach rather high abundances in the upper layers, although with a sizable uncertainty of around one order of magnitude. In the case of GJ\,1214b, the transmission spectrum observed with JWST appears rather featureless, indicative of optically thick clouds or hazes, although there is some tentative evidence of CO$_2$ and CH$_4$ \citep{Schlawin2024a}. The combined analysis of HST and JWST transmission spectra favors an atmosphere with an extremely rich metallicity where hydrogen is no longer the main constituent \citep{Ohno2025}. According to our calculations, CO$_2$ is naturally expected at high metallicities, but CH$_4$ is not predicted to be that abundant unless the C/O ratio increases above unity. In our model of GJ\,1214b, where we  adopted a metallicity of 1000\,$\times$\,solar and a solar C/O ratio of 0.55, the uncertainties associated with the predicted abundances of the major observable molecules, such as H$_2$O, CO, CO$_2$, and H$_2$S, are rather small (see Fig.\,\ref{fig:warm_neptunes}). Therefore, a disequilibrium model can be faithfully used to constrain the metallicity and C/O ratio, although in GJ\,436b and GJ\,1214b this is hampered by the presence of clouds or hazes. A more favorable case is that of the warm Neptune GJ\,3470b, where there is evidence for H$_2$O, CO$_2$, CH$_4$, and SO$_2$ in the JWST transmission spectrum \citep{Beatty2024}. The detection of SO$_2$, a species clearly formed in disequilibrium by photochemistry, is consistent with our calculations for GJ\,436b, where the adopted metallicity of 100\,$\times$\,solar is the same inferred for GJ\,3470b. As discussed before, the uncertainty in the calculated abundance of SO$_2$ due to (photo)chemical kinetics and vertical mixing is non-negligible.

\subsection{Most critical reactions} \label{subsec:critical_reactions}

\begin{table*}
\small
\caption{The 50 most critical reactions according to their global Spearman correlation coefficient.}
\label{table:reactions}
\centering
\begin{tabular}{lcrrrrrrr}
\hline \hline
Reaction $i$& \multicolumn{1}{c}{Error} & \multicolumn{1}{c}{$\sum_{p=1}^{6} C_p^i$} & \multicolumn{6}{c}{$C^i$} \\
\cline{4-9}
& & & \multicolumn{1}{c}{WASP-33b} & \multicolumn{1}{c}{HD\,209458b} & \multicolumn{1}{c}{HD\,189733b} & \multicolumn{1}{c}{WASP-39b} & \multicolumn{1}{c}{GJ\,436b} & \multicolumn{1}{c}{GJ\,1214b} \\
\hline

\hline
\\
\multicolumn{9}{c}{Chemical reactions}\\
\hline
    S + H$_2$ + M $\rightarrow$ H$_2$S + M                    & C &  60.65 &     -- &   6.30 &  19.67 &  17.69 &  14.65 &   2.34 \\
    H + H + M $\rightarrow$ H$_2$ + M                         & A &  57.37 &  10.43 &  18.37 &   5.88 &  22.70 &     -- &     -- \\
    S + OH $\rightarrow$ SO + H                               & B &  29.50 &   0.31 &   0.88 &   2.99 &  15.76 &   7.30 &   2.27 \\
    NS + NH$_2$ $\rightarrow$ H$_2$S + N$_2$                  & C &  27.63 &     -- &   7.16 &   1.47 &   9.78 &   4.62 &   4.61 \\
    Si + O + M $\rightarrow$ SiO + M                          & C &  15.40 &  12.33 &     -- &     -- &   3.07 &     -- &     -- \\
    NH + N + M $\rightarrow$ N$_2$H + M                       & C &  14.88 &  14.54 &     -- &     -- &   0.34 &     -- &     -- \\
    P + PH $\rightarrow$ P$_2$ + H                            & B &  11.49 &     -- &   0.42 &   2.14 &     -- &   4.24 &   4.69 \\
    N$_2$H$_3$ + M $\rightarrow$ N$_2$H$_2$ + H + M           & C &  10.63 &     -- &   2.71 &   3.40 &   1.94 &   2.58 &     -- \\
    N + NH$_3$ $\rightarrow$ N$_2$H + H$_2$                   & C &   9.92 &     -- &   5.35 &   2.30 &   2.27 &     -- &     -- \\
    SiS + O $\rightarrow$ SiO + S                             & B &   9.08 &     -- &   5.04 &   0.39 &   3.65 &     -- &     -- \\
    S + SH $\rightarrow$ S$_2$ + H                            & B &   8.73 &     -- &   0.73 &   2.42 &   0.45 &   3.25 &   1.88 \\
    PH + H$_2$ + M $\rightarrow$ PH$_3$ + M                   & C &   7.90 &     -- &     -- &     -- &     -- &   4.40 &   3.50 \\
    S$_2$ + H + M $\rightarrow$ HS$_2$ + M                    & C &   7.83 &     -- &     -- &     -- &     -- &   7.11 &   0.72 \\
    Si + S + M $\rightarrow$ SiS + M                          & C &   7.53 &   5.36 &     -- &     -- &   2.17 &     -- &     -- \\
    OH + H$_2$ $\rightarrow$ H + H$_2$O                       & A &   6.61 &     -- &     -- &   3.95 &     -- &   2.66 &     -- \\
    Si + CO $\rightarrow$ SiO + C                             & B &   6.28 &   5.79 &     -- &   0.50 &     -- &     -- &     -- \\
    S + CH$_3$ $\rightarrow$ H$_2$CS + H                      & C &   6.16 &     -- &   0.50 &   2.69 &     -- &   1.17 &   1.80 \\
    OH + CO $\rightarrow$ CO$_2$ + H                          & A &   5.82 &     -- &   0.38 &   0.64 &     -- &   2.11 &   2.69 \\
    P + PO$_2$ $\rightarrow$ PO + PO                          & C &   5.19 &     -- &     -- &     -- &     -- &   2.95 &   2.23 \\
    CH$_3$O + CO $\rightarrow$ CO$_2$ + CH$_3$                & A &   5.12 &     -- &     -- &     -- &     -- &     -- &   5.12 \\
    HNC + OH $\rightarrow$ HNCO + H                           & B &   5.04 &     -- &     -- &   2.34 &     -- &   1.17 &   1.53 \\
    H$_2$S$_2$ + H $\rightarrow$ H$_2$S + SH                  & B &   4.56 &     -- &     -- &     -- &     -- &   1.99 &   2.57 \\
    HCS + H $\rightarrow$ CS + H$_2$                          & C &   4.54 &     -- &     -- &     -- &     -- &   2.12 &   2.41 \\
    PO + N $\rightarrow$ PN + O                               & B &   4.48 &     -- &   0.32 &   1.24 &   0.45 &   1.42 &   1.05 \\
    SiO + O + M $\rightarrow$ SiO$_2$ + M                     & C &   4.44 &   0.95 &   0.92 &   1.17 &   1.41 &     -- &     -- \\
    CS + OH $\rightarrow$ OCS + H                             & B &   4.44 &     -- &   0.39 &   2.39 &     -- &   0.83 &   0.83 \\
    HCP + O $\rightarrow$ PH + CO                             & C &   4.33 &     -- &   0.66 &   0.75 &   0.92 &   1.00 &   1.00 \\
    Si + H$_2$S $\rightarrow$ SiS + H$_2$                     & B &   4.28 &     -- &   2.00 &   2.27 &     -- &     -- &     -- \\
    S + OCS $\rightarrow$ S$_2$ + CO                          & A &   3.67 &     -- &     -- &     -- &     -- &   0.39 &   3.28 \\
    Ti + O + M $\rightarrow$ TiO + M                          & C &   3.63 &   1.91 &   0.83 &     -- &   0.89 &     -- &     -- \\
    TiO + O + M $\rightarrow$ TiO$_2$ + M                     & C &   3.54 &   0.79 &   0.78 &   0.78 &   1.19 &     -- &     -- \\
    TiO + OH $\rightarrow$ TiO$_2$ + H                        & B &   3.52 &   0.50 &   0.50 &   1.48 &   1.04 &     -- &     -- \\
    Si + CO$_2$ $\rightarrow$ SiO + CO                        & B &   3.46 &     -- &     -- &   3.46 &     -- &     -- &     -- \\
    P + OH $\rightarrow$ PO + H                               & B &   3.45 &     -- &   1.27 &   1.80 &   0.38 &     -- &     -- \\
    S + NH $\rightarrow$ NS + H                               & C &   3.28 &     -- &     -- &     -- &     -- &   1.04 &   2.24 \\
\hline
\\
\multicolumn{9}{c}{Photoreactions}\\
\hline
    H$_2$O + h$\nu$ $\rightarrow$ OH + H                      & A &  55.37 &     -- &  20.40 &  16.51 &     -- &   8.82 &   9.63 \\
    ~~~~~~~~~~~~~~~~~$\rightarrow$ O($^1$D) + H$_2$           & A &  55.37 &     -- &  20.40 &  16.51 &     -- &   8.82 &   9.63 \\
    ~~~~~~~~~~~~~~~~~$\rightarrow$ O + H + H                  & A &  55.37 &     -- &  20.40 &  16.51 &     -- &   8.82 &   9.63 \\
    S$_2$ + h$\nu$ $\rightarrow$ S + S                        & A &  42.07 &     -- &   3.83 &   2.81 &   6.34 &   6.43 &  22.66 \\
    SH + h$\nu$ $\rightarrow$ S + H                           & A &  31.02 &   0.31 &  17.16 &   2.22 &  11.33 &     -- &     -- \\
    SO + h$\nu$ $\rightarrow$ S + O                           & A &  22.02 &     -- &   0.42 &   0.38 &  20.34 &   0.40 &   0.48 \\
    P$_2$ + h$\nu$ $\rightarrow$ P + P                        & C &  16.61 &     -- &   1.89 &   3.77 &   0.36 &   5.36 &   5.23 \\
    PH$_2$ + h$\nu$ $\rightarrow$ PH + H                      & C &  16.22 &     -- &   8.15 &   8.07 &     -- &     -- &     -- \\
    SiS + h$\nu$ $\rightarrow$ Si + S                         & C &  15.69 &   0.70 &   6.84 &   6.69 &   1.46 &     -- &     -- \\
    CH + h$\nu$ $\rightarrow$ C + H                           & B &  13.79 &  13.49 &   0.30 &     -- &     -- &     -- &     -- \\
    TiO + h$\nu$ $\rightarrow$ Ti + O                         & C &  12.73 &   1.60 &   3.32 &   3.62 &   4.19 &     -- &     -- \\
    PN + h$\nu$ $\rightarrow$ P + N                           & C &  11.47 &   0.67 &   2.99 &   3.75 &   1.48 &   2.21 &   0.37 \\
    SiH + h$\nu$ $\rightarrow$ Si + H                         & B &  10.78 &  10.21 &   0.57 &     -- &     -- &     -- &     -- \\
    OH + h$\nu$ $\rightarrow$ O + H                           & A &   9.99 &     -- &   9.99 &     -- &     -- &     -- &     -- \\
    ~~~~~~~~~~~~~~~$\rightarrow$ O($^1$D) + H                 & A &   9.99 &     -- &   9.99 &     -- &     -- &     -- &     -- \\
    SiO + h$\nu$ $\rightarrow$ Si + O                         & B &   9.79 &     -- &   3.41 &   6.02 &   0.36 &     -- &     -- \\
    PO + h$\nu$ $\rightarrow$ P + O                           & C &   3.89 &     -- &   1.36 &   0.86 &   0.93 &   0.40 &   0.33 \\
    TiO$_2$ + h$\nu$ $\rightarrow$ TiO + O                    & C &   3.83 &     -- &   0.90 &   1.43 &   1.50 &     -- &     -- \\
\hline
\end{tabular}
\tablenoteb{\\
Note.-- The term $\sum_{p=1}^{6} C_p^i$ stands for the sum of $C^i$ over the six planets, where $C^i$ is the global correlation coefficient of reaction $i$, see Eq.\,(\ref{eq:p}).
}
\end{table*}

There are some reactions that have a big impact on the chemical composition, while others have a moderate or little influence. The sensitivity analysis allows us to identify the most influential reactions using correlation coefficients between reaction rate coefficients and species abundances. Pearson correlation coefficients capture correlations when the relationship between reaction rate perturbations and abundance changes is linear, while Spearman captures monotonic relationships although not necessarily linear. Given that chemical networks are highly nonlinear, we give preference here to Spearman correlation coefficients. We nevertheless note that for the particular cases investigated here, the Pearson and Spearman approaches yield similar coefficients and the conclusions on the most influential reactions remain the same no matter whether Pearson or Spearman is used. In Table\,\ref{table:reactions} we list the 50 most critical reactions, as given by the largest sum of global Spearman correlation coefficients over the six planets modeled. We  grouped the processes into two categories: chemical reactions and photoreactions. In general, the list of most influential reactions is composed of a few processes with a low uncertainty (A) and many with poorly constrained rate coefficients or cross sections (labeled B and C).

\begin{figure}
\centering
\includegraphics[angle=0,width=\columnwidth]{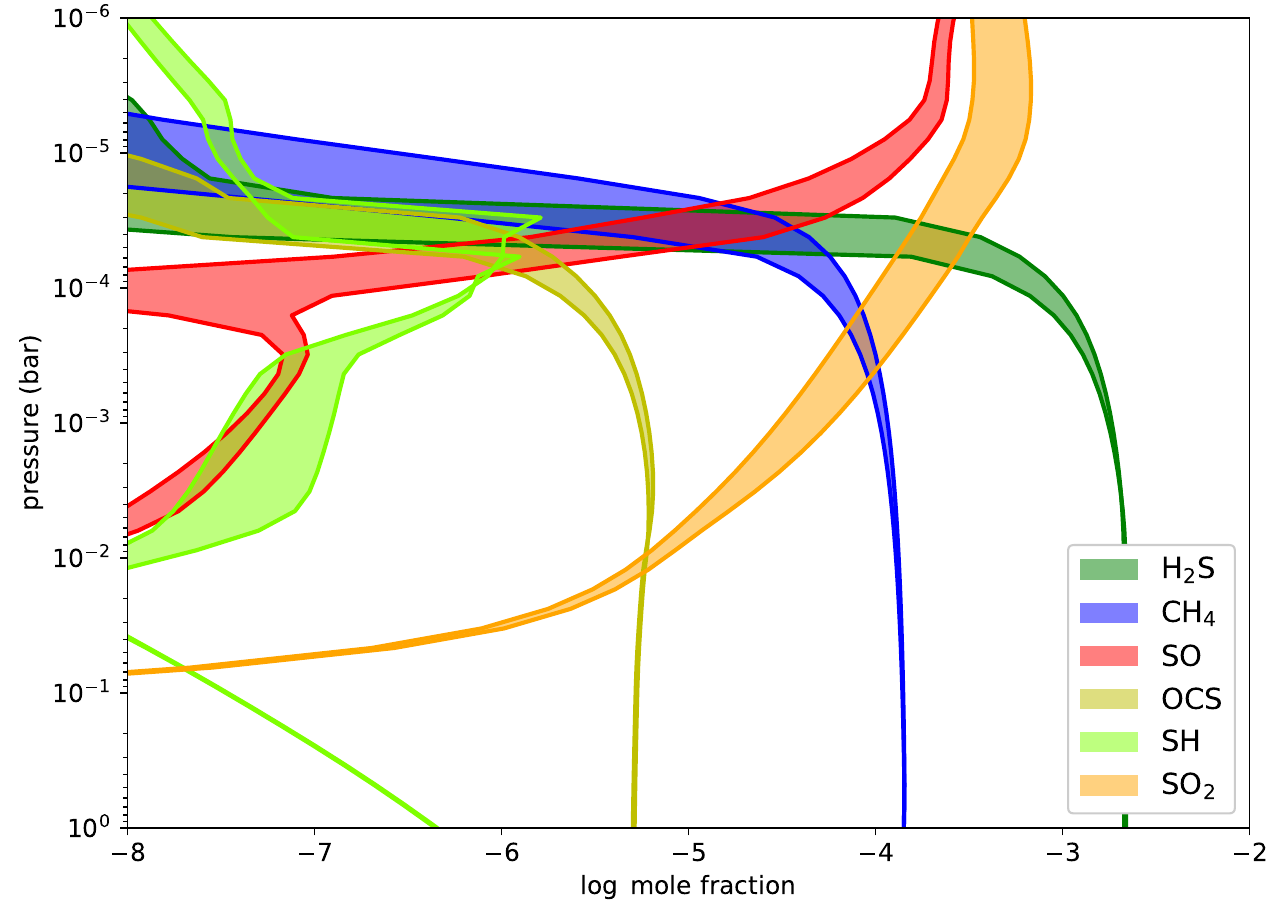}
\caption{Impact of the reaction S + H$_2$ + M $\rightarrow$ H$_2$S + M on the abundances of selected species in GJ\,436b. The shaded areas show the variation in the calculated abundances when the adopted rate coefficient is ten times below and above the nominal value.} \label{fig:s+h2_gj436b}
\end{figure}

In the category of chemical reactions, the three-body recombination of two H atoms emerges as a critical reaction in ultra-hot and hot Jupiters, while the reactions OH + H$_2$ $\rightarrow$ H + H$_2$O and OH + CO $\rightarrow$ CO$_2$ + H also have  a high multi-planet global correlation coefficient. These reactions have low uncertainties because they have been extensively studied experimentally at different temperatures \citep{Jacobs1965,Frost1993,Lissianski1995,Baulch2005,Varga2016}, and thus there is little room for a significant improvement in the accuracy. There are however many other influential reactions whose rate coefficients are poorly known and which are ideal targets for an improved characterization. It is worth noting that most of them involve sulfur, phosphorus, silicon, and titanium, which illustrates that the chemistry of these elements is not as well known as that  of carbon, nitrogen, and oxygen. The three-body recombination of S and H$_2$ has the largest multi-planet global correlation coefficient, with an impact on the chemistry of sulfur in all planets except WASP-33b. The kinetics of this reaction is rather uncertain \citep{Zahnle2016}. The high-pressure limit rate coefficient is merely a guess, while the low-pressure limit is based on the reverse process, on which there are conflicting predictions \citep{Shiina1998,Karan2004,Raj2020}. It would therefore be interesting to revisit this reaction experimentally or theoretically. This reaction affects sulfur species, but also other major species such as H, OH, CH$_4$, and NH$_3$. To illustrate how the uncertainty of this sole reaction impacts the abundances of various key species, in Fig.\,\ref{fig:s+h2_gj436b} we show the change in the abundance of selected species caused by varying the reaction rate coefficient within its uncertainty. It is seen that the assigned uncertainty of one order of magnitude results in abundance variations of a factor of a few. Other reactions amenable to be further studied are S + OH $\rightarrow$ SO + H, which has only been studied at room temperature \citep{Burkholder2020} and has a large impact on the chemistry of sulfur in WASP-39b, NS + NH$_2$ $\rightarrow$ H$_2$S + N$_2$, which is thought to be barrierless based on calculations by \cite{Nguyen1998}, P + PH $\rightarrow$ P$_2$ + H, which controls much of the chemistry of phosphorus in the warm Neptunes GJ\,436b and GJ\,1214b and for which there are inconclusive rate coefficient calculations \citep{Kaye1983,Lee2024}, and the reaction N + NH$_3$ $\rightarrow$ N$_2$H + H$_2$, in which case no detailed calculations or experiments are available \citep{Moses2016}. There are also several three-body recombinations, such as Si + O, NH + N, N$_2$H$_2$ + H, PH + H$_2$, S$_2$ + H, and Si + S, that would be very interesting to study. All of them have large uncertainties and are highly influential reactions.

Several photodissociation processes are also highlighted in the ranking of most influential reactions, which is not surprising given that the largest abundance dispersion occurs in the upper atmospheric layers due to photochemistry (see Fig.\,\ref{fig:hot_jupiters} and Fig.\,\ref{fig:warm_neptunes}). In the top position, the photodissociation of water reaches the largest multi-planet global correlation coefficient among all photoreactions, even if this process has a low uncertainty. The photoabsorption cross section of H$_2$O has been well characterized over a wide wavelength range \citep{Chan1993,Fillion2004,Mota2005}, although photodissociation yields are only known at certain discrete wavelengths \citep{Slanger1982,Crovisier1989}. Given the remarkable influence of this process, a more precise characterization of the yields of the various fragmentation channels as a function of wavelength would be of high interest. Other photodissociation processes with large multi-planet global correlation coefficients and low uncertainties are those of S$_2$, SH, SO, and OH, all of them with relatively well-known cross section data \citep{Phillips1981,Nee1986,Hrodmarsson2023}. There are also several critical but poorly known photodissociation processes, such as those of P$_2$, PH$_2$, SiS, CH, TiO, PN, SiH, SiO, PO, and TiO$_2$, that would benefit from a more thorough investigation. As in the case of chemical reactions, most of them involve  S-, P-, Si-, and Ti-bearing molecules, the photodissociation of which is not accurately known.

The identification of the most critical reactions listed in Table\,\ref{table:reactions} relies on multi-planet global correlation coefficients. This means that these are the reactions with the largest overall influence on different abundant species present in different types of planets, covering ultra-hot Jupiters, hot Jupiters, and warm Neptunes. It is possible to restrict the identification of the most critical reactions to a specific species in a given layer of a particular planet. As discussed in Sect.\,\ref{subsec:abundance_uncertainties}, there are several species, such as HCN, SO$_2$, PH$_3$, and TiO, that have important uncertainties due to (photo)chemical kinetics in certain atmospheric layers of certain planets. For example, HCN shows an important abundance uncertainty above the 10$^{-3}$ bar pressure level in HD\,189733b (see Fig.\,\ref{fig:hot_jupiters}). The sensitivity analysis indicates that the reaction HNC + OH $\rightarrow$ HNCO + H is the one with the largest correlation, with a Spearman coefficient of $-$0.67. That is, the faster the reaction the lower the abundance of HCN. This is different from the scheme elucidated by \cite{Moses2011}, where the reaction CH$_3$NH$_2$ + H $\rightarrow$ CH$_2$NH$_2$ + H$_2$ was identified as the rate-limiting step in the mechanism of conversion of CH$_4$ and NH$_3$ into HCN. The different findings may be in part due to the fact that the reaction between CH$_3$NH$_2$ and H is included in our network with a low uncertainty (type A; \citealt{Kerkeni2007}), which makes it less influential, while that of HNC and OH has a larger uncertainty (type B) based on conflicting results from theoretical calculations \citep{Lin1992,Dean2000,Bunkan2004}. SO$_2$ is another species with a sizable abundance uncertainty in the upper atmosphere of WASP-39b (see Fig.\,\ref{fig:hot_jupiters}) and GJ\,436b (see Fig.\,\ref{fig:warm_neptunes}). The reaction with the largest correlation with SO$_2$ is S + OH $\rightarrow$ SO + H, which is positively correlated with a Spearman coefficient between $+$0.49 and $+$0.61. This reaction also appears  as a key step in the formation scheme of SO$_2$ elucidated by \cite{Tsai2023}. In GJ\,436b, the reaction S + H$_2$ + M $\rightarrow$ H$_2$S + M, which is at the top of the most critical reactions listed in Table\,\ref{table:reactions}, also appears  correlated with SO$_2$, in this case negatively, with a Spearman coefficient of $-$0.38. The phosphorus hydride PH$_3$ is also predicted to have an important abundance uncertainty above the 10$^{-2}$ bar pressure level in GJ\,1214b (see Fig.\,\ref{fig:warm_neptunes}), where the most influential reaction becomes PH + H$_2$ + M $\rightarrow$ PH$_3$ + M, which has a Spearman correlation coefficient of $+$0.60 and a large uncertainty in its rate coefficient (type C). Finally, TiO is also predicted to have a rather uncertain abundance in the upper layers of the hot Jupiters HD\,209458b and HD\,189733b (see Fig.\,\ref{fig:hot_jupiters}). In this case the main process responsible for this uncertainty is the photodissociation of TiO (see \citealt{Agundez2025}), which has a Spearman correlation coefficient of $-$0.96 and whose cross section is not known (type C). Dedicated studies on the aforementioned reactions would  reduce the uncertainties in the calculated abundances of HCN, SO$_2$, PH$_3$, and TiO in the exoplanet atmospheres where they are relevant constituents.

The method used here to identify the most influential reactions takes into account the uncertainty in the kinetics of the reaction, and thus it is biased toward processes with large uncertainties. In other words, reactions with large uncertainties are more likely to be influential than those with small uncertainties. This approach is used on purpose to identify which  processes  affect  the uncertainties in the abundances most strongly, and whose detailed inspection and more accurate determination of the rate coefficient or cross section would serve to reduce the uncertainty of the chemical model. The formalism is different from that used to identify which is the rate-limiting step in key conversions happening in exoplanet atmospheres, such as CH$_4$\,$\rightarrow$\,CO and NH$_3$\,$\rightarrow$\,N$_2$ \citep{Moses2011,Tsai2017}, where reaction uncertainties are not taken into account. In any case, it is interesting to see which reactions arise as the most important ones in the two approaches. For example, in HD\,189733b methane quenches at $\sim$1 bar (see Fig.\,\ref{fig:hot_jupiters}), where the reactions showing the largest correlation with CH$_4$ are OH + CH$_3$ $\rightarrow$ CH$_2$OH + H and OH + CH$_3$ + M $\rightarrow$ CH$_3$OH + M, both with a Spearman coefficient of $-$0.45. Interestingly, these two reactions, which arise as the two most influential in spite of their low assigned uncertainty (type A based on \citealt{Jasper2007}), are also the rate-limiting steps, one or another depending on the pressure and eddy diffusion, in the CH$_4$\,$\rightarrow$\,CO conversion \citep{Moses2011,Tsai2017}. In the case of the NH$_3$\,$\rightarrow$\,N$_2$ conversion, \cite{Moses2011} identify the rate-limiting step as N$_2$H$_3$ + M $\rightarrow$ N$_2$H$_2$ + H + M in HD\,189733b, while we found that this reaction is also the most correlated  with NH$_3$ at 10 bar, where NH$_3$ quenches, with a Spearman coefficient of $-$0.68. Unlike in the CH$_4$\,$\rightarrow$\,CO conversion, in this case the kinetics of the most critical reaction is poorly constrained (type C; \citealt{Konnov2001,Hwang2003}) and thus would be a nice target for a more thorough investigation.

\section{Conclusions} \label{sec:conclusions}

We carried out a sensitivity analysis to quantify the abundance uncertainties resulting from a state-of-the-art chemical model of an exoplanet atmosphere. Concretely, we studied how the uncertainties in reaction rate coefficients, photodissociation cross sections, and vertical mixing strength propagate to abundances in chemical models of the gaseous giants WASP-33b, HD\,209458b, HD\,189733b, WASP-39b, GJ\,436b, and GJ\,1214b.

In general, we found that (i) abundance uncertainties are low within the observable atmosphere (1-10$^{-6}$ bar), below one order of magnitude, and in many cases even below a factor of two; (ii) vertical mixing is found to dominate the abundance uncertainties at a similar or even higher level than (photo)chemical kinetics; (iii) abundance uncertainties are smaller in planets with a composition close to chemical equilibrium than in those dominated by photochemistry; (iv) molecules such as H$_2$O, CO, CO$_2$, and SiO show low abundance uncertainties, and thus can be confidently compared with observations, while others such as HCN, SO$_2$, PH$_3$, and TiO have more uncertain abundances, and are therefore amenable for a better characterization of their (photo)chemistry.

We also identified the (photo)reactions that become most critical for the establishment of the global atmospheric composition. The chemical reactions H + H + M $\rightarrow$ H$_2$ + M, OH + H$_2$ $\rightarrow$ H + H$_2$O, and OH + CO $\rightarrow$ CO$_2$ + H, and the photodissociations of H$_2$O, S$_2$, SH, SO, and OH arise as influential processes, although all them have low uncertainties, and thus there is little room for  improvement in the accuracy. Other critical processes that are poorly known comprise the three-body associations S + H$_2$, Si + O, NH + N, N$_2$H$_2$ + H, PH + H$_2$, and S$_2$ + H, the chemical reactions S + OH $\rightarrow$ SO + H, NS + NH$_2$ $\rightarrow$ H$_2$S + N$_2$, P + PH $\rightarrow$ P$_2$ + H, and N + NH$_3$ $\rightarrow$ N$_2$H + H$_2$, and the photodissociations of P$_2$, PH$_2$, SiS, CH, TiO, PN, SiH, SiO, PO, and TiO$_2$. A better characterization of these latter processes, many of which involve S-, P-, Si-, and Ti-bearing species, through either experiments or calculations should lead to more accurate predicted abundances.

\begin{acknowledgements}

We acknowledge funding support from Spanish Ministerio de Ciencia, Innovaci\'on y Universidades through grant PID2023-147545NB-I00 and the computational resources provided by the DRAGO computer cluster managed by SGAI-CSIC, and the Galician Supercomputing Center (CESGA). We thank A. Lira-Barria for useful discussions. We thank the referees for a critical reading and for very useful comments that helped to improve this manuscript.

\end{acknowledgements}


\begin{thebibliography}{}

\bibitem[Ag\'undez et al.(2014a)]{Agundez2014a} Ag\'undez, M., Parmentier, V., Venot, O., et al. 2014a, \aap, 564, A73
\bibitem[Ag\'undez et al.(2014b)]{Agundez2014b} Ag\'undez, M., Venot, O., Selsis, F., \& Iro, N. 2014b, \apj, 781, 68
\bibitem[Ag\'undez(2025)]{Agundez2025} Ag\'undez, M. 2025, \aap, 699, A306
\bibitem[Alderson et al.(2023)]{Alderson2023} Alderson, L., Wakeford, H. R., Alam, M. K., et al. 2023, \nature, 614, 664
\bibitem[Balmer et al.(2025)]{Balmer2025} Balmer, W. O., Kammerer, J., Pueyo, L., et al. 2025, \aj, 169, 209
\bibitem[Baulch et al.(2005)]{Baulch2005} Baulch, D. L., Bowman, C. T., Cobos, C. J., et al. 2005, \jpcrd, 34, 757
\bibitem[Beatty et al.(2024)]{Beatty2024} Beatty, T. G., Welbanks, L., Schlawin, E., et al. 2024, \apj, 970, L10
\bibitem[Behr et al.(2023)]{Behr2023} Behr, P. R., France, K., Brown, A., et al. 2023, \aj, 166, 35
\bibitem[Bell et al.(2024)]{Bell2024} Bell, T. J., Crouzet, N., Cubillos, P. E., et al. 2024, \natastro, 8, 879
\bibitem[Benne et al.(2022)]{Benne2022} Benne, B., Dobrijevic, M., Cavali\'e, T., et al. 2022, \aap, 667, A169
\bibitem[Blain et al.(2024a)]{Blain2024a} Blain, D., Landman, R., Molli{\`e}re, P., \& Dittmann, J. 2024a, \aap, 690, A63
\bibitem[Blain et al.(2024b)]{Blain2024b} Blain, D., S\'anchez-L\'opez, A., \& Molli{\`e}re, P. 2024b, \aj, 167, 179
\bibitem[Boucher et al.(2021)]{Boucher2021} Boucher, A., Darveau-Bernier, A., Pelletier, S., et al. 2021, \aj, 162, 233
\bibitem[Boyajian et al.(2015)]{Boyajian2015} Boyajian, T., von Braun, K., Feiden, G. A., et al. 2015, \mnras, 447, 846
\bibitem[Bunkan et al.(2004)]{Bunkan2004} Bunkan, A. J. C., Tang, Y., Sellev\r{a}g, S. R., \& Nielsen, C. J. 2004, \jpca, 118, 5279
\bibitem[Burkholder et al.(2020)]{Burkholder2020} Burkholder, J. B., Sander, S. P., Abbatt, J. P. D., et al. 2020, JPL Publication 19-5
\bibitem[Byrne et al.(2024)]{Byrne2024} Byrne, A. N., Xue, C., Van Voorhis, T., \& McGuire, B. A. 2024, \pccp, 26, 26734
\bibitem[Carrasco et al.(2008)]{Carrasco2008} Carrasco, N., Alcaraz, C., Dutuit, O., et al. 2008, \pss, 56, 1644
\bibitem[Carter et al.(2024)]{Carter2024} Carter, A. L., May, E. M., Espinoza, N., et al. 2024, \natastro, 8, 1008
\bibitem[Castelli \& Kurucz(2004)]{Castelli2004} Castelli, F. \& Kurucz, R. L. 2004, arXiv:\texttt{astro-ph/0405087}
\bibitem[Casasayas-Barris et al.(2021)]{Casasayas-Barris2021} Casasayas-Barris, N., Palle, E., Stangret, M., et al. 2021, \aap, 647, A26
\bibitem[Chan et al.(1993)]{Chan1993} Chan, W. F., Cooper, G., \& Brion, C. E. 1993, \cp, 178, 387
\bibitem[Cloutier et al.(2021)]{Cloutier2021} Cloutier, R., Charbonneau, D., Deming, D., et al. 2021, \aj, 162, 174
\bibitem[Collier Cameron et al.(2010)]{CollierCameron2010} Collier Cameron, A., Guenther, E., Smalley, B., et al. 2010, \mnras, 407, 507
\bibitem[Cont et al.(2021)]{Cont2021} Cont, D., Yan, F., Reiners, A., et al. 2021, \aap, 651, A33
\bibitem[Cont et al.(2022)]{Cont2022} Cont, D., Yan, F., Reiners, A., et al. 2022, \aap, 668, A53
\bibitem[Crovisier(1989)]{Crovisier1989} Crovisier, J. 1989, \aap, 213, 459
\bibitem[Dang et al.(2025)]{Dang2025} Dang, L., Bell, T. J., Shu, Y. (Z.), et al. 2025, \aj, 169, 32
\bibitem[Dean \& Bozzelli(2000)]{Dean2000} Dean, A. M. \& Bozzelli, J. W. 2000, Gas-Phase Combustion Chemistry, p 125, Springer
\bibitem[D\'esert et al.(2008)]{Desert2008} D\'esert, J.-M., Vidal-Madjar, A., Lecavelier des Etangs, A., et al. 2008, \aap, 492, 585
\bibitem[Dobrijevic \& Parisot(1998)]{Dobrijevic1998} Dobrijevic, M. \& Parisot, M. 1998, \pss, 46, 491
\bibitem[Dobrijevic et al.(2003)]{Dobrijevic2003} Dobrijevic, M., Ollivier, J. L., Billebaud, F., et al. 2003, \aap, 398, 335
\bibitem[Dobrijevic et al.(2010a)]{Dobrijevic2010a} Dobrijevic, M., H\'ebrard, E., Plessis, S., et al. 2010a, \asr, 45, 77
\bibitem[Dobrijevic et al.(2010b)]{Dobrijevic2010b} Dobrijevic, M., Cavali\'e, T., H\'ebrard, E., et al. 2010b, \pss, 58, 1555
\bibitem[Evans-Soma et al.(2025)]{Evans-Soma2025} Evans-Soma, T. M., Sing, D. K., Barstow, J. K., et al. 2025, \natastro, 9, 845
\bibitem[Faedi et al.(2011)]{Faedi2011} Faedi, F., Barros, S. C. C., Anderson, D. R., et al. \aap, 531, A40
\bibitem[Fillion et al.(2004)]{Fillion2004} Fillion, J.-H., Ruiz, J., Yang, X.-F., et al. 2004, \jcp, 120, 6531
\bibitem[Finnerty et al.(2024)]{Finnerty2024} Finnerty, L., Xuan, J. W., Xin, Y., et al. 2024, \aj, 167, 43
\bibitem[France et al.(2016)]{France2016} France, K., Loyd, R. O. P., Youngblood, A., et al. 2016, \apj, 820, 89
\bibitem[Frost et al.(1993)]{Frost1993} Frost, M. J., Sharkey, P., \& Smith, I. W. M. 1993, \jpc, 97, 12254
\bibitem[Fu et al.(2024)]{Fu2024} Fu, G., Welbanks, L., Deming, D., et al. 2024, \nature, 632, 752
\bibitem[Gapp et al.(2025)]{Gapp2025} Gapp, C., Evans-Soma, T. M., Barstow, J. K., et al. \aj, 169, 341
\bibitem[Giacobbe et al.(2021)]{Giacobbe2021} Giacobbe, P., Brogi, M., Gandhi, S., et al. 2021, \nature, 592, 205
\bibitem[Ginzburg \& Sari(2015)]{Ginzburg2015} Ginzburg, S. \& Sari, R. 2015, \apj, 803, 111
\bibitem[Goos et al.(2018)]{Goos2018} Goos, E., Burcat, A., \& Ruscic, B. 2018, Extended Third Millenium Ideal Gas Thermochemical Database with updates from Active Thermochemical Tables, \url{https://burcat.technion.ac.il/}
\bibitem[Hardy et al.(2015)]{Hardy2015} Hardy, L. K., Butterley, T., Dhillon, V. S., et al. 2015, \mnras, 454, 4316
\bibitem[Harps{\o}e et al.(2013)]{Harpsoe2013} Harps{\o}e, K. B. W., Hardis, S., Hinse, T. C., et al. 2013, \aap, 549, A10
\bibitem[Hawker et al.(2018)]{Hawker2018} Hawker, G. A., Madhusudhan, N., Cabot, S. H. C., \& Gandhi, S. 2018, \apj, 863, L11
\bibitem[Haynes et al.(2015)]{Haynes2015} Haynes, K., Mandell, A. M., Madhusudhan, N., et al. 2015, ApJ, 806, 146
\bibitem[Heays et al.(2017)]{Heays2017} Heays, A. N., Bosman, A. D., \& van Dishoeck, E. F. 2017, \aap, 602, A105
\bibitem[H\'ebrard et al.(2006)]{Hebrard2006} H\'ebrard, E., Dobrijevic, M., B\'enilan, Y., \& Raulin, F. 2006, \jppc, 7, 211
\bibitem[H\'ebrard et al.(2007)]{Hebrard2007} H\'ebrard, E., Dobrijevic, M., B\'enilan, Y., \& Raulin, F. 2007, \pss, 55, 1470
\bibitem[H\'ebrard et al.(2009)]{Hebrard2009} H\'ebrard, E., Dobrijevic, M., Pernot, P., et al. 2009, \jpca, 113, 11227
\bibitem[Hrodmarsson \& van Dishoeck(2023)]{Hrodmarsson2023} Hrodmarsson, H. R. \& van Dishoeck, E. F. 2023, \aap, 675, A25
\bibitem[Huebner et al.(1992)]{Huebner1992} Huebner, W. F., Keady, J. J., \& Lyon, S. P. 1992, Ap\&SS, 195, 1
\bibitem[Hwang \& Mebel(2003)]{Hwang2003} Hwang, D.-Y. \& Mebel, A. M. 2003, \jpca, 107, 2865
\bibitem[Inglis et al.(2024)]{Inglis2024} Inglis, J., Batalha, N. E., Lewis, N. K., et al. 2024, \apj, 973, L41
\bibitem[Jacobs et al.(1965)]{Jacobs1965} Jacobs, T. A., Giedt, R. R., \& Cohen, N. 1965, \jcp, 43, 3688
\bibitem[Jasper et al.(2007)]{Jasper2007} Jasper, A. W., Klippenstein, S. J., Harding, L. B., \& Ruscic, B. 2007, \jpca, 111, 3932
\bibitem[JWST Transiting Exoplanet Community Early Release Science Team(2023)]{JWST2023} JWST Transiting Exoplanet Community Early Release Science Team (Ahrer, E.-M. et al.) 2023, \nature, 614, 649
\bibitem[Karan et al.(2004)]{Karan2004} Karan, K., Mehrotra, A. K., \& Behie, L. A. 2004, AlChE J., 45, 383
\bibitem[Kawashima \& Min(2021)]{Kawashima2021} Kawashima, Y. \& Min, M. 2021, \aap, 656, A90
\bibitem[Kaye \& Strobel(1983)]{Kaye1983} Kaye, J. A. \& Strobel, D. F. 1983, \grs, 10, 957
\bibitem[Keller-Rudek et al.(2013)]{Keller-Rudek2013} Kaller-Rudek, H., Moortgat, G. K., Sander, R., \& S\"orensen, R. 2013, Earth Syst. Sci. Data, 5, 365
\bibitem[Kempton et al.(2023)]{Kempton2023} Kempton, E. M.-R., Zhang, M., Bean, J. L. et al. 2023, \nature, 620, 67
\bibitem[Kerkeni \& Clary(2007)]{Kerkeni2007} Kerkeni, B. \& Clary, D. C. 2007, \cpl, 438, 1
\bibitem[Klein et al.(2024)]{Klein2024} Klein, B., Debras, F., Donati, J.-F., et al. 2024, \mnras, 527, 544
\bibitem[Knutson et al.(2014)]{Knutson2014} Knutson, H. A., Benneke, B., Deming, D., \& Homeier, D. 2014, \nature, 505, 66
\bibitem[Koll et al.(2022)]{Koll2022} Koll, D. D. B. 2022, \apj, 924, 134
\bibitem[Komazek et al.(2022)]{Komacek2022} Komacek, T. D., Gao, P., Thorngren, D. P., et al. 2022, \apj, 941, L40
\bibitem[Konnov \& De Ruyck(2001)]{Konnov2001} Konnov, A. A. \& De Ruyck, J. 2001, \cflame, 124, 106
\bibitem[Lee et al.(2024)]{Lee2024} Lee, E. K. H., Tsai, S.-M., Moses, J. I., et al. 2024, \apj, 976, 231
\bibitem[Lehmann et al.(2015)]{Lehmann2015} Lehmann, H., Guenther, E., Sebastian, D., et al. 2015, \aap, 578, L4
\bibitem[Lin et al.(1992)]{Lin1992} Lin, M. C., He, Y., \& Melius, C. F. 1992, \ijck, 24, 489
\bibitem[Line et al.(2011)]{Line2011} Line, M. R., Vasisht, G., Chen, P., et al. 2011, \apj, 738, 32
\bibitem[Line et al.(2014)]{Line2014} Line, M. R., Knutson, H., Wolf, A. S., \& Yung, Y. L. 2014, \apj, 783, 70
\bibitem[Line et al.(2016)]{Line2016} Line, M. R., Stevenson, K. B., Bean, J., et al. 2016, \aj, 152, 203
\bibitem[Lira-Barria(2025)]{Lira-Barria2025} Lira-Barria, A. 2025, PhD Thesis, KU Leuven
\bibitem[Lissianski et al.(1995)]{Lissianski1995} Lissianski, V., Yang, H., Qin, Z., et al. 1995, Chem. Phys. Lett., 240, 57
\bibitem[Lothringer et al.(2018)]{Lothringer2018} Lothringer, J. D., Benneke, B., Crossfield, I. J. M., et al. 2018, \aj, 155, 66
\bibitem[Lothringer et al.(2022)]{Lothringer2022} Lothringer, J. D., Sing, D. K., Rustamkulov, Z., et al. 2022, \nature, 604, 49
\bibitem[Loyd et al.(2016)]{Loyd2016} Loyd, R. O. P., France, K., Youngblood, A., et al. 2016, \apj, 824, 102
\bibitem[Madhusudhan \& Seager(2009)]{Madhusudhan2009} Madhusudhan, N. \& Seager, S. 2009, \apj, 707, 24
\bibitem[Madhusudhan \& Seager(2011)]{Madhusudhan2011} Madhusudhan, N. \& Seager, S. 2011, \apj, 729, 41
\bibitem[Madhusudhan(2012)]{Madhusudhan2012} Madhusudhan, N. 2012, \apj, 758, 36
\bibitem[McBride et al.(2002)]{McBride2002} McBride, B. J., Zehe, M. J., \& Gordon, S. 2002, NASA Technical Publication TP-2002-211556
\bibitem[Melo et al.(2024)]{Melo2024} Melo, E., Souto, D., Cunha, K., et al. 2024, \apj, 973, 90
\bibitem[Molli{\`e}re et al.(2015)]{Molliere2015} Molli{\`e}re, P., van Boekel, R., Dullemond, C., et al. 2015, \apj, 813, 47
\bibitem[Moses et al.(2011)]{Moses2011} Moses, J. I., Visscher, C., Fortney, J. J., et al. 2011, \apj, 737, 15
\bibitem[Moses et al.(2013a)]{Moses2013a} Moses, J. I., Madhusudhan, N., Visscher, C., \& Freedman, R. S. 2013a, \apj, 763, 25
\bibitem[Moses et al.(2013b)]{Moses2013b} Moses, J. I., Line, M. R., Visscher, C., et al. 2013b, \apj, 777, 34
\bibitem[Moses et al.(2016)]{Moses2016} Moses, J. I., Marley, M. S., Zahnle, K., et al. 2016, \apj, 829, 66
\bibitem[Moses et al.(2022)]{Moses2022} Moses, J. I., Tremblin, P., Venot, O., \& Miguel, Y. 2022, \expas, 53, 279
\bibitem[Mota et al.(2005)]{Mota2005} Mota, R., Parafita, R., Giuliani, A., et al. 2005, \cpl, 416, 152
\bibitem[Mraz et al.(2024)]{Mraz2024} Mraz, G., Darveau-Bernier, A., Boucher, A., et al. 2024, \apj, 975, L42
\bibitem[Mukherjee et al.(2025)]{Mukherjee2025} Mukherjee, S., Schlawin, E., Bell, T. J., et al. 2025, \apj, 982, L39
\bibitem[Nahar(2020)]{Nahar2020} Nahar, S. N. 2020, Atoms, 8, 68
\bibitem[Nahar \& Hinojosa-Aguirre(2024)]{Nahar2024} Nahar, S. N. \& Hinojosa-Aguirre, G. 2024, Atoms, 12, 22
\bibitem[Nee \& Lee(1986)]{Nee1986} Nee, J. B \& Lee, L. C. 1986, \jcp, 84, 5303
\bibitem[Nguyen et al.(1998)]{Nguyen1998} Nguyen, L. T., Le, T. N., \& Nguyen, M. T. 1998, \jcsft, 94, 3541
\bibitem[Nugroho et al.(2021)]{Nugroho2021} Nugroho, S. K., Kawahara, H., Gibson, N. P., et al. 2021, \apj, 910, L9
\bibitem[Ohno et al.(2025)]{Ohno2025} Ohno, K., Schlawin, E., Bell, T. J., et al. 2025, \apj, 979, L7
\bibitem[Palle et al.(2025)]{Palle2025} Palle, E., Biazzo, K., Bolmont, E., et al. 2025, \expas, 59, 29
\bibitem[Paredes et al.(2021)]{Paredes2021} Paredes, L. A., Henry, T. J., Quinn, S. N., et al. 2021, \aj, 162, 176
\bibitem[Parmentier et al.(2013)]{Parmentier2013} Parmentier, V., Showman, A. P., \& Lian, Y. 2013, \aap, 558, A91
\bibitem[Phillips(1981)]{Phillips1981} Phillips, L. F. 1981, \jpc, 85, 3994
\bibitem[Pinhas et al.(2019)]{Pinhas2019} Pinhas, A., Madhusudhan, N., Gandhi, S., \& MacDonald, R. 2019, \mnras, 482, 1485
\bibitem[Powell et al.(2024)]{Powell2024} Powell, D., Feinstein, A. D., Lee, E. K. H., et al. 2024, \nature, 626, 979
\bibitem[Press et al.(1992)]{Press1992} Press, W. H., Flannery, B. P., Teukolsky, S. A., \& Vetterling, W. T. 1992, Numerical Recipes in Fortran 77: The Art of Scientific computing, Cambridge Univ. Press, 1992
\bibitem[Raj et al.(2020)]{Raj2020} Raj, A., Ibrahim, S., \& Jagannath, A. 2020, Progr. Energy Comb. Sci., 80, 100848
\bibitem[Rustamkulov et al.(2023)]{Rustamkulov2023} Rustamkulov, Z., Sing, D. K., Mukherjee, S., et al. 2023, \nature, 614, 659
\bibitem[Schlawin et al.(2024a)]{Schlawin2024a} Schlawin, E., Ohno, K., Bell, T. J., et al. 2024a, \apj, 974, L33
\bibitem[Schlawin et al.(2024b)]{Schlawin2024b} Schlawin, E., Mukherjee, S., Ohno, K., et al. 2024b, \aj, 168, 104
\bibitem[Shiina et al.(1998)]{Shiina1998} Shiina, H., Miyoshi, A., \& Matsui, H. 1998, \jpca, 102, 3556
\bibitem[Sing et al.(2016)]{Sing2016} Sing, D. K., Fortney, J. J., Nikolov, N., et al. 2016, \nature, 529, 59
\bibitem[Sing et al.(2024)]{Sing2024} Sing, D. K., Rustamkulov, Z., Thorngren, D. P., et al. 2024, \nature, 630, 831
\bibitem[Slanger \& Black(1982)]{Slanger1982} Slanger, T. G. \& Black, G. 1982, \jcp, 77, 2432
\bibitem[Smith et al.(2011)]{Smith2011} Smith, A. M. S., Anderson, D. R., Skillen, I., et al. 2011, \mnras, 416, 2096
\bibitem[Southworth(2010)]{Southworth2010} Southworth, J. 2010, \mnras, 408, 1689
\bibitem[Stevenson et al.(2010)]{Stevenson2010} Stevenson, K. B., Harrington, J., Nymeyer, S., et al. 2010, \nature, 464, 1161
\bibitem[Thorngren et al.(2019)]{Thorngren2019} Thorngren, D., Gao, P., \& Fortney, J. J. 2019, \apj, 884, L6
\bibitem[Torres(2007)]{Torres2007} Torres, G. 2007, \apj, 671, L65
\bibitem[Tsai et al.(2017)]{Tsai2017} Tsai, S.-M., Lyons, J. R., Grosheintz, L., et al. 2017, \apjs, 228, 20
\bibitem[Tsai et al.(2021)]{Tsai2021} Tsai, S.-M., Malik, M., Kitzmann, D., et al. 2021, \apj, 923, 264
\bibitem[Tsai et al.(2023)]{Tsai2023} Tsai, S.-M., Lee, E. K. H., Powell, D., et al. 2023, \nature, 617, 483
\bibitem[van Sluijs et al.(2023)]{vanSluijs2023} van sluijs, L., Birkby, J. L., Lothringer, J., et al. 2023, \mnras, 522, 2145
\bibitem[Varga et al.(2016)]{Varga2016} Varga, T., Olm, C., Nagy, T., et al. 2016, \ijck, 48, 407
\bibitem[Vasyunin et al.(2004)]{Vasyunin2004} Vasyunin, A. I., Sobolev, A. M., Wiebe, D. S., \& Semenov, D. A. 2004, \astronl, 30, 566
\bibitem[Vasyunin et al.(2008)]{Vasyunin2008} Vasyunin, A. I., Semenov, D., Henning, Th., et al. 2008, \apj, 672, 629
\bibitem[Venot et al.(2012)]{Venot2012} Venot, O., H\'ebrard, E., Ag\'undez, M., et al. 2012, \aap, 546, A43
\bibitem[Venot et al.(2014)]{Venot2014} Venot, O., Ag\'undez, M., Selsis, F., et al. 2014, \aap, 562, A51
\bibitem[Venot et al.(2018)]{Venot2018} Venot, O., Drummond, B., Miguel, Y., et al. 2018, \expas, 46, 101
\bibitem[Venot et al.(2019)]{Venot2019} Venot, O., Bounaceur, R., Dobrijevic, M., et al. 2019, \aap, 624, A58
\bibitem[Venot et al.(2020)]{Venot2020} Venot, O., Cavali\'e, T., Bounaceur, R., et al. 2020, \aap, 634, A78
\bibitem[Wakelam et al.(2005)]{Wakelam2005} Wakelam, V., Selsis, F., Herbst, E., \& Caselli, P. 2005, \aap, 444, 883
\bibitem[Wakelam et al.(2006)]{Wakelam2006} Wakelam, V., Herbst, E., \& Selsis, F. 2006, \aap, 451, 551
\bibitem[Wakelam et al.(2010a)]{Wakelam2010a} Wakelam, V., Herbst, E., Le Bourlot, J., et al. 2010a, \aap, 517, A21
\bibitem[Wakelam et al.(2010b)]{Wakelam2010b} Wakelam, V., Smith, I. W. M., Herbst, E., et al. 2010b, \ssr, 156, 13
\bibitem[Welbanks et al.(2024)]{Welbanks2024} Welbanks, L., Bell, T. J., Beatty, T. G., et al. 2024, \nature, 630, 836
\bibitem[Woods et al.(2009)]{Woods2009} Woods, T. N., Chamberlin, P. C., Harder, J. W., et al. 2009, \grl, 36, L01101
\bibitem[Xue et al.(2024)]{Xue2024} Xue, Q., Bean, J. L., Zhang, M., et al. 2024, \apj, 963, L5
\bibitem[Yan et al.(2022)]{Yan2022} Yan, F., Pall\'e, E., Reiners, A., et al. 2022, \aap, 661, L6
\bibitem[Yang et al.(2024)]{Yang2024} Yang, Y., Chen, G., Yan, F., et al. 2024, \apj, 971, L8
\bibitem[Youngblood et al.(2016)]{Youngblood2016} Youngblood, A., France, K., Loyd, R. O. P., et al. 2016, \apj, 824, 101
\bibitem[Zahnle et al.(2009)]{Zahnle2009} Zahnle, K., Marley, M. S., Freedman, R. S., et al. 2009, \apj, 701, L20
\bibitem[Zahnle et al.(2016)]{Zahnle2016} Zahnle, K., Marley, M. S., Morley, C. V., \& Moses, J. I. 2016, \apj, 824, 137
\bibitem[Zhang et al.(2018)]{Zhang2018} Zhang, M., Knutson, H. A., Kataria, T., et al. 2018, \aj, 155, 83
\bibitem[Zhang et al.(2025)]{Zhang2025} Zhang, M., Paragas, K., Bean, J. L., et al. 2025, \aj, 169, 38

\end{thebibliography}
\end{document}